\documentclass[longauth]{aa}

\usepackage{graphicx}
\usepackage{txfonts}
\usepackage{amsmath}
\usepackage{adjustbox} 
\usepackage{caption} 
\usepackage[para,online,flushleft]{threeparttable} 
\usepackage{rotating}
\usepackage{booktabs} 
\usepackage{url} 
\usepackage[table]{xcolor} 
\usepackage{bm}
\usepackage[flushleft]{threeparttable}

\usepackage{longtable}
\usepackage{pdflscape}
\usepackage{gensymb}
\usepackage{natbib}

\usepackage[
hidelinks,
colorlinks=true,
linkcolor=blue,
citecolor=blue]{hyperref}

\begin{document}

   \title{The ALMA survey to Resolve exoKuiper belt Substructures (ARKS)}

   \subtitle{VI. Asymmetries and offsets}

\author{J.~B.~Lovell\textsuperscript{1}\fnmsep\thanks{E-mail: joshualovellastro@gmail.com} \and A.~S.~Hales\textsuperscript{2,3,4} \and G.~M.~Kennedy\textsuperscript{5,6} \and S.~Marino\textsuperscript{7} \and J.~Olofsson\textsuperscript{8} \and A.~M.~Hughes\textsuperscript{9} \and E.~Mansell\textsuperscript{9} \and B.~C.~Matthews\textsuperscript{10,11} \and T.~D.~Pearce\textsuperscript{6} \and A.~A.~Sefilian\textsuperscript{12} \and D.~J.~Wilner\textsuperscript{1} \and B.~Zawadzki\textsuperscript{9} \and M.~Booth\textsuperscript{13} \and M.~Bonduelle\textsuperscript{14} \and A.~Brennan\textsuperscript{15} \and C.~del~Burgo\textsuperscript{16,17} \and J.~M.~Carpenter\textsuperscript{3} \and G.~Cataldi\textsuperscript{18,19} \and E.~Chiang\textsuperscript{20} \and A.~Fehr\textsuperscript{1} \and Y.~Han\textsuperscript{21} \and Th.~Henning\textsuperscript{22} \and A.~V.~Krivov\textsuperscript{23} \and P.~Luppe\textsuperscript{15} \and J.~P.~Marshall\textsuperscript{24} \and S.~Mac~Manamon\textsuperscript{15} \and J.~Milli\textsuperscript{14} \and A.~Mo\'or\textsuperscript{25} \and M.~C.~Wyatt\textsuperscript{26} \and S.~Ertel\textsuperscript{12,27} \and M.~R.~Jankovic\textsuperscript{28} \and \'A.~K\'osp\'al\textsuperscript{25,29,22} \and M.~A.~MacGregor\textsuperscript{30} \and L.~Matr\`a\textsuperscript{15} \and S.~P\'erez\textsuperscript{31,4,32} \and P.~Weber\textsuperscript{31,4,32}} 

\institute{
Center for Astrophysics | Harvard \& Smithsonian, 60 Garden St, Cambridge, MA 02138, USA \and
National Radio Astronomy Observatory, 520 Edgemont Road, Charlottesville, VA 22903-2475, United States of America \and
Joint ALMA Observatory, Avenida Alonso de C\'ordova 3107, Vitacura 7630355, Santiago, Chile \and
Millennium Nucleus on Young Exoplanets and their Moons (YEMS), Chile \and
Malaghan Institute of Medical Research, Gate 7, Victoria University, Kelburn Parade, Wellington, New Zealand \and
Department of Physics, University of Warwick, Gibbet Hill Road, Coventry CV4 7AL, UK \and
Department of Physics and Astronomy, University of Exeter, Stocker Road, Exeter EX4 4QL, UK \and
European Southern Observatory, Karl-Schwarzschild-Strasse 2, 85748 Garching bei M\"unchen, Germany \and
Department of Astronomy, Van Vleck Observatory, Wesleyan University, 96 Foss Hill Dr., Middletown, CT, 06459, USA \and
Herzberg Astronomy \& Astrophysics, National Research Council of Canada, 5071 West Saanich Road, Victoria, BC, V9E 2E9, Canada \and
Department of Physics \& Astronomy, University of Victoria, 3800 Finnerty Rd, Victoria, BC V8P 5C2, Canada \and
Department of Astronomy and Steward Observatory, The University of Arizona, 933 North Cherry Ave, Tucson, AZ, 85721, USA \and
UK Astronomy Technology Centre, Royal Observatory Edinburgh, Blackford Hill, Edinburgh EH9 3HJ, UK \and
Univ. Grenoble Alpes, CNRS, IPAG, F-38000 Grenoble, France \and
School of Physics, Trinity College Dublin, the University of Dublin, College Green, Dublin 2, Ireland \and
Instituto de Astrof\'isica de Canarias, Vía L\'actea S/N, La Laguna, E-38200, Tenerife, Spain \and
Departamento de Astrof\'isica, Universidad de La Laguna, La Laguna, E-38200, Tenerife, Spain \and
National Astronomical Observatory of Japan, Osawa 2-21-1, Mitaka, Tokyo 181-8588, Japan \and
Department of Astronomy, Graduate School of Science, The University of Tokyo, Tokyo 113-0033, Japan \and
Department of Astronomy, University of California, Berkeley, Berkeley, CA 94720-3411, USA \and
Division of Geological and Planetary Sciences, California Institute of Technology, 1200 E. California Blvd., Pasadena, CA 91125, USA \and
Max-Planck-Insitut f\"ur Astronomie, K\"onigstuhl 17, 69117 Heidelberg, Germany \and
Astrophysikalisches Institut und Universit\"atssternwarte, Friedrich-Schiller-Universit\"at Jena, Schillerg\"a{\ss}chen 2-3, 07745 Jena, Germany \and
Academia Sinica Institute of Astronomy and Astrophysics, 11F of AS/NTU Astronomy-Mathematics Building, No.1, Sect. 4, Roosevelt Rd, Taipei 106319, Taiwan. \and
Konkoly Observatory, HUN-REN Research Centre for Astronomy and Earth Sciences, MTA Centre of Excellence, Konkoly-Thege Mikl\'os \'ut 15-17, 1121 Budapest, Hungary \and
Institute of Astronomy, University of Cambridge, Madingley Road, Cambridge CB3 0HA, UK \and
Large Binocular Telescope Observatory, The University of Arizona, 933 North Cherry Ave, Tucson, AZ, 85721, USA \and
Institute of Physics Belgrade, University of Belgrade, Pregrevica 118, 11080 Belgrade, Serbia \and
Institute of Physics and Astronomy, ELTE E\"otv\"os Lor\'and University, P\'azm\'any P\'eter s\'et\'any 1/A, 1117 Budapest, Hungary \and
Department of Physics and Astronomy, Johns Hopkins University, 3400 N Charles Street, Baltimore, MD 21218, USA \and
Departamento de Física, Universidad de Santiago de Chile, Av. V\'ictor Jara 3493, Santiago, Chile \and
Center for Interdisciplinary Research in Astrophysics Space Exploration (CIRAS), Universidad de Santiago, Chile
}

   \date{Received July 23, 2025; accepted November 13, 2025}

  \abstract
   {Asymmetries in debris discs provide unique clues to understand the evolution and architecture of planetary systems. 
   Previous studies of debris discs at (sub)millimetre wavelengths have suggested the presence of asymmetries in a wide variety of systems, yet the lack of sufficiently sensitive high-resolution observations means that the typical properties of debris disc asymmetries have not been studied at the population level. 
   The aim of the ALMA survey to Resolve exoKuiper belt Substructures (ARKS) is to expand our understanding of radial and vertical dust structures, as well as gas distributions and kinematics, in debris discs. The ARKS sample of 24 highly resolved targets in ALMA's Bands 6 and 7 (1.1--1.4\,mm and 0.8--1.1\,mm, respectively) provided a unique opportunity to study their asymmetries.}
   {Here, in ARKS~VI, we present a systematic analysis of the asymmetries and stellocentric offsets present in the ALMA continuum data for the ARKS survey.
   Our aims are to i) identify asymmetries in debris disc dust distributions, ii) quantify debris disc asymmetry properties, and iii) discuss the potential origins of debris disc asymmetries.
   This work is the first systematic analysis of asymmetries in a large sample of well-resolved discs at (sub)millimetre wavelengths.}
   {We utilised empirical methods to identify emission asymmetries (relative to disc major and minor axes, and azimuthal disc locations) and the presence of offset emission between disc centres and the locations of the host stars, via an analysis of their calibration procedures and disc properties. 
   We associated observational asymmetry types (offset, major and/or minor axis, azimuthal) and plausible physical classes (arcs, eccentricities, and possible clumps and warps) associated with each source.}
   {We show that there are ten systems, almost half of the ARKS sample, that host  either a continuum emission asymmetry or offset emission.
   Three systems host offsets (HD~15115, HD~32297, and HD~109573 (HR~4796)), four host azimuthal asymmetries (HD~9672 (49~Ceti), HD~92945, HD~107146, and HD~121617), two host an asymmetry in their major axis (HD~10647 (q$^1$~Eri), and HD~39060 ($\beta$~Pic)), and one hosts an asymmetry in their minor axis (HD~61005).
   We attribute the offset asymmetries to non-zero eccentricities, and three of the azimuthal asymmetries to arcs. 
   The presence of an asymmetry or offset in the ARKS sample appears to be correlated with the fractional luminosity of cold dust.
   We tentatively suggest that continuum asymmetries are more prevalent in CO-rich debris discs, suggesting that gas interactions may drive debris dust asymmetries. 
   We identify seven other tentative asymmetries, including four in  distinct ARKS systems and three in systems with otherwise significant asymmetries.}
   {This study demonstrates that debris disc asymmetries in the ARKS sample are common, and plausibly so in the wider population of debris discs at (sub)-millimetre wavelengths.  
   This means that (sub)-millimetre asymmetries plausibly await discovery in debris discs as we probe these with higher sensitivity and resolution.
   Throughout, we highlight future studies to further investigate the origins of debris disc asymmetries, and build on the work presented here. }


   \maketitle

\section{Introduction}
\label{sec:intro}
Debris discs are dusty rings around stars that are sustained by continual planetesimal belt collisions over megayear to gigayear timescales \citep{Wyatt08, Matthews14, Hughes18, Marino2022}.
As mature circumstellar discs, debris discs provide critical insights into the evolution of planetary systems following the dispersal of their protoplanetary discs, through the stellar main sequence, and beyond to post-main sequence planetary systems \citep[see e.g.][]{Bonsor10, Wyatt15, Farihi16}.
Improvements in sensitivity and angular resolution over recent years at  (sub)millimetre wavelengths -- in particular, with the James Clerk Maxwell Telescope (JCMT), Submillimeter Array (SMA), and Atacama Large Millimeter/submillimeter Array (ALMA) -- have shown the diversity of debris disc structures and substructures. 
This diversity can now provide a greater understanding as to how planetary systems evolve, via studies of their continuum dust and carbon monoxide (CO) gas morphologies \citep[see e.g. the surveys of][and references therein]{Holland17, Moor2017, Lovell21a, MatraReasons25}.

Asymmetries and offsets have been strongly associated with some of the best-studied debris discs, particularly in scattered light images. 
The warp in the $\beta$ Pictoris disc \citep{Mouillet+1997, Heap00}, later identified as due to a second disc component \citep{Kalas05,Golimowski06}, is one of the earliest and most well-known examples. 
The offset of the Fomalhaut disc centre from the stellar position is obvious at all wavelengths at which the star is detectable \citep[e.g.][]{Acke12, Su13,  MacGregor17,Gaspar23, Chittidi+2025, Lovell+25_Fom}. 
Asymmetries that vary over time have even been detected in the AU Mic disc \citep{Boccaletti15}, with various potential explanations \citep{Sezestre17, Chiang17, Grady20, Thebault18}.
Though these asymmetries are among the most well-known and pronounced, many discs  exhibit asymmetries of some kind.
\cite{Crotts24} conducted an empirical analysis of 26 debris discs,  many of them less well-studied, imaged with the Gemini Planet Imager (GPI) and found that the majority of imaged discs (${>}70$\%) have one or more discernibly asymmetric features, such as brightness, radial extent, or {\it JHK} colours, as well as warps and azimuthal features. 
However, studies have not yet found comparably high occurrence rates of debris disc asymmetries at longer wavelengths.

Asymmetries in debris discs at millimetre wavelengths have been reported in data prior to ALMA, for example from the Submillimeter Array  \citep[SMA; see e.g.][in the case of HD\,15115]{Macgregor+2015} and the \textit{James Clerk Maxwell} Telescope \citep[JCMT; see e.g.][in the case of Fomalhaut, and other targets in the SONS -- \hbox{SCUBA-2} Observations of Nearby Stars -- sample]{Holland+2003, Panic+2013, Holland17}.
Moreover, several millimetre asymmetric structures in debris discs were known before ARKS, including the CO gas clump in the south of the $\beta$~Pic (HD~39060) disc \citep[e.g.][]{Dent2014}, the continuum (dust) emission clump in the main ring of q$^1$~Eri \citep[HD~10647;][]{Lovell21c}, and the eccentric discs of Fomalhaut (HD~216956), HD~53143, and HD~202628 \citep[e.g.][]{MacGregor17, Faramaz2019, Kennedy20, Macgregor22}.
In other cases, pre-ARKS ALMA data have suggested hints of asymmetric emission that have not been definitively proven, for example in the case of the ARKS target HD~92945 \citep[][]{Marino19} and $\varepsilon$~Eri \citep[HD~22049; see][]{Booth2023}.
In some cases, asymmetric clumps have been found in systems that were then found to originate from galactic or extra-galactic confusion \citep[e.g. in the case of background sources being co-located in the discs of HD~95086, Fomalhaut and HD~22049; see][]{Booth19, KennedyLovell2023, Booth2023} highlighting the need for long-term monitoring of asymmetries to verify their origins.
Moreover, asymmetries have been detected in a range of systems in scattered light images without counterpart asymmetries in (sub)millimetre thermal emission images \citep[see e.g.][]{Olofsson2019, Olofsson+2022}, which may result from either observational differences (i.e. differences in their signal-to-noise ratio (S/N) or resolution) or physical differences (with different wavelength observations being more or less sensitive to distinct physical features that trace underlying asymmetric structures), highlighting the need for multi-wavelength monitoring to quantify their origins and behaviour.

Asymmetries in circumstellar discs more generally contain vital clues   to the processes driving their evolution.
These are present at early evolutionary stages in protoplanetary discs \citep[see e.g.][]{Andrews20} and at later evolutionary stages in debris discs \citep[see also][]{Hughes18, Marino2022}. Arcs and crescents are the rarest type of substructure in protoplanetary discs (less common than rings or spirals) based on their occurrence rates \citep{Bae+2023}, and are detected in up to 30\% of massive discs, falling to around 5\% in less massive discs.

Asymmetries in debris discs have been uniquely linked to a range of dynamically induced substructures.
A commonly invoked scenario is one in which these asymmetries can be driven directly by planet--disc interactions \citep[see e.g.][]{Dong2020}.
Examples include scenarios in which eccentric planets drive eccentricities \citep[e.g.][]{Quillen06, Pearce14, Kennedy20, LynchLovell21, LovellLynch2023} or spirals into discs \citep[][]{Hahn2003, Wyatt05, Pearce15, Rodet2017, Sefilian+2021, Sefilian+2023}.
Circular but migrating planets, on the other hand, can trap planetesimals in mean-motion resonances, which can produce clumps in discs \citep[e.g.][]{Wyatt2003, Wyatt06, Reche08, Booth2023, Friebe+2022}.
Planets can also carve gaps in discs beyond their own orbital radii via secular resonances \citep[e.g.][]{Pearce15, Yelverton+2018, Sefilian+2021, Sefilian+2023}.
Orbital inclination misalignments between planets and discs can also cause vertical warps in the discs \citep[see e.g.][]{Mouillet+1997, Augeraeu+2001, Sefilian+2025}.
However, planets are not uniquely responsible for inducing asymmetries in discs. Other scenarios may also drive these features, for instance stellar flybys \citep[][]{Reche08, Rodet2017, Pfalzner+2024}, stellar binary interactions \citep{Farhat+2023}, gas-dust interactions \citep{Lyra2013}, inheritance from earlier stages of evolution \citep{Nealon+2018, Kennedy20}, slow modes of self-gravitating discs \citep{Jalali2012}, or via giant impacts between large planetesimals in the disc \citep{Jackson14, Dent14, Jones+2023}.
Quantifying asymmetric emission features therefore allows one to understand their origins, and details the histories of planetary systems.

Debris discs show wavelength-dependent architectures. 
Whilst larger millimetre- to centimetre-sized grains typically trace the orbits of their parent planetesimals, smaller (sub)micron-sized grains are more strongly affected by radiation forces, which can drag material inwards, for example via the Poynting--Robertson effect, or blow small grains out of discs via radiation pressure \citep{Matthews14, Hughes18}.
The physics that dictate how grains are transported in debris discs can alter their observed asymmetric morphologies as a function of wavelength, for example with scattered light observations being able to observe dust out to much larger radii (i.e. where ISM interactions may plausibly shape their morphologies) and with millimetre observations being more sensitive to asymmetries inherent in planetesimal belts \citep[see e.g.][where planetary system dynamics are more likely to dominate their morphologies]{Wyatt06, Lee16, LynchLovell21, Jones+2023}.
These distinctions mean it is important to study asymmetries from a multi-wavelength perspective to understand the origins of asymmetric debris disc structures.

The ALMA survey to Resolve exoKuiper belt Substructures (ARKS) is the first ALMA large programme  dedicated to studying the structures and dynamics of a sample of the nearest and brightest debris discs, providing unprecedented sensitivity and resolution, sufficient to resolve their radial and vertical morphologies.
Most pertinent to the analysis and discussion in this paper are papers ARKS~I \citep{overview_arks}, which discusses the overview, reduction, and calibration of the sample; ARKS~II \citep{rad_arks}, which models and presents (axisymmetric) disc radial profiles; ARKS~III \citep{ver_arks}, which models and presents (axisymmetric) disc vertical profiles; and ARKS~V \citep{scat_arks}, which presents SPHERE scattered light images and profiles of discs. 
In this paper we present a systematic analysis of the sample's continuum asymmetries and offsets.
We performed most of our analysis empirically and solely on dust continuum emission products, as provided by the work presented in ARKS~I, and further detailed in Appendix~\ref{sec:AppendixSources}.
Assessment of the CO gas asymmetries will be presented in a future ARKS paper.

The rest of this paper is structured as follows. In \S\ref{sec:offsetsAsymms} we present our methods for identifying dust continuum asymmetries or offsets. In \S\ref{sec:results} we discuss our findings for each source, and then in \S\ref{sec:discussion} in terms of the population.
We summarise our findings and conclude in \S\ref{sec:conclusions}. Technical details are presented in Appendices \ref{sec:AppendixSources}--\ref{sec:app_selfsubmaps}.

\begin{figure*}
    \centering
    \includegraphics[clip, trim={0cm 0cm 0cm 0cm}, width=0.9
    \linewidth]{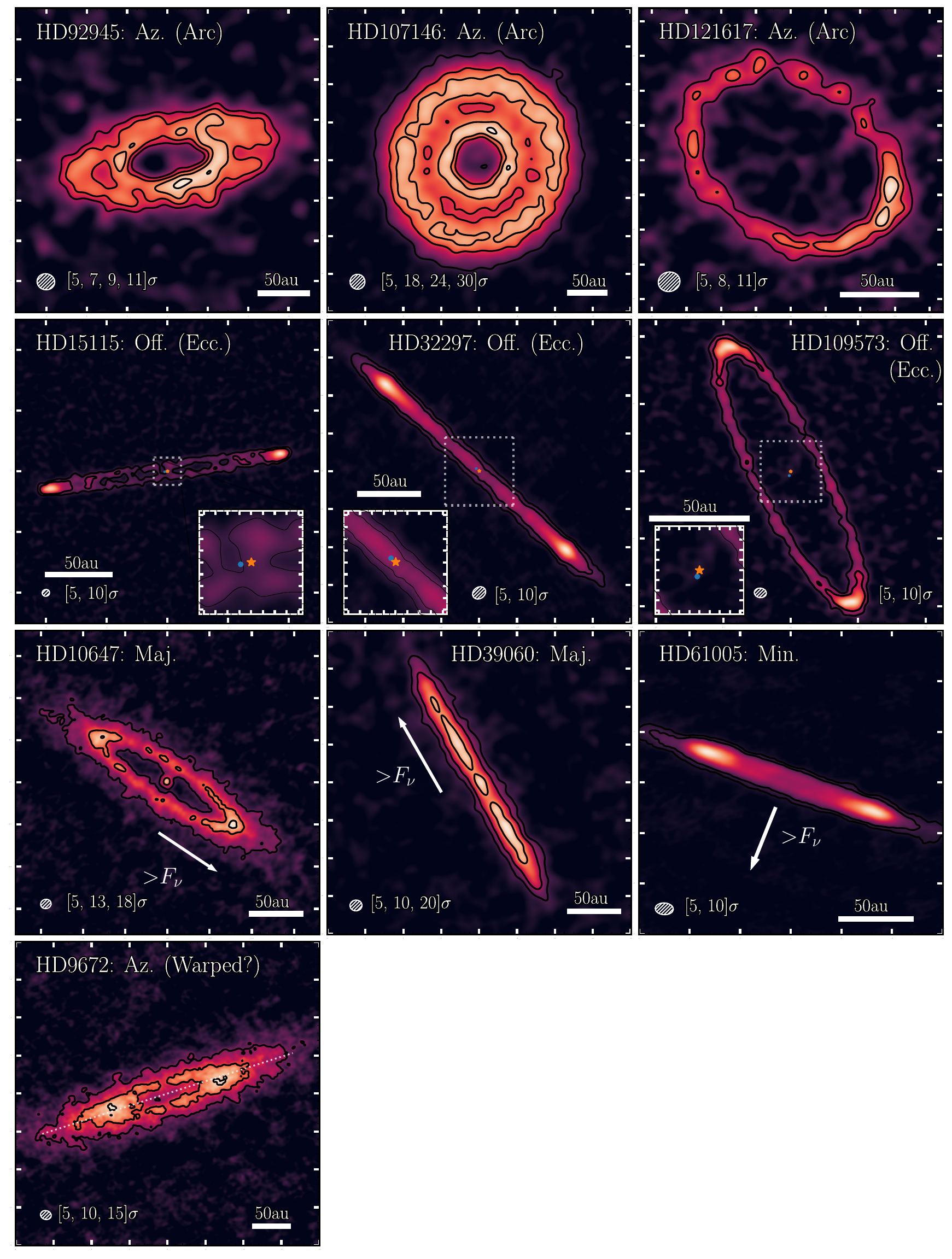}
    \caption{Gallery of the ten ARKS systems with evidence of asymmetric or offset emission, and their respective observational asymmetry type. The arrows indicate systems that host a major or minor axis with the directionality pointing towards the brighter side. All systems are shown with different contour levels to accentuate their asymmetric features. We present the synthesised beams in the lower left of each panel, as well as 50\,au scale bars in the lower right. For systems with significant offsets, we show insets zoomed into their central regions, denoting their phase (stellar) centres as orange stars and best-fit disc centres as blue dots. In all panels north is up and east is left. From left to right, then top to bottom, the images are cropped to fields of view that in both axes are ${\pm}7.5''$, ${\pm}8''$, ${\pm}0.875''$, ${\pm}2.5''$, ${\pm}1''$, ${\pm}1.125''$, ${\pm}9''$, ${\pm}8''$, ${\pm}3''$, and ${\pm}4''$, respectively.}
    \label{fig:gallery}
\end{figure*}

\section{Stellocentric offsets and emission asymmetries}
There are principally four types of observational asymmetry that we investigated for with these ARKS data.
We searched for observational asymmetries based on these presenting asymmetric emission about their i) major axes (Maj), ii) minor axes (Min), iii) at some localised azimuthal location (Az), or iv) if these systems are host to a stellocentric offset (Off), i.e. a global offset between the disc emission centre and its central star.
These asymmetry types are noted beside each relevant system in Table~\ref{tab:asymmetries}, each of which we conclude on per system in \S\ref{sec:results}.
Whilst asymmetric emission and/or offsets may be present in any given system, disc geometries can challenge our ability to definitively interpret their physical origin, thus in some cases we cannot ascribe to these a physical classification.
In this section we discuss the methodologies by which we assess ARKS data to be host to either a stellocentric offset or asymmetric emission.
The data we analysed is described in Appendix~\ref{sec:AppendixSources}.

\label{sec:offsetsAsymms}
\subsection{Offset assessments}
\label{sec:offsets}
We assessed all 24 ARKS targets for the presence of stellocentric offsets by analysing which discs have Gaia DR3 stellar locations significantly offset from their best-fit disc centre positions. 
These checks were made based on each target's fitted $\Delta$RA and $\Delta$Decl. values that were measured consistently for the full ARKS sample using symmetric Gaussian radial distribution models, as presented in ARKS~I \citep[see their Table~B.1][]{overview_arks}.

In ARKS~I \citep{overview_arks}, HD~15115, HD~32297, HD~39060, HD~95086, and HD~109573 all have best-fit $\Delta$RA and/or $\Delta$Decl. offsets that are significantly offset from 0 (by more than $3\sigma$).
Since HD~95086's data suffers from the undersubtraction of a background source \citep[as noted in][]{overview_arks}, we deem this disc centre offset too biased to conclude that it is real. 
In the case of HD~39060, we attributed the disc centre offset to the data having a systematically poorer astrometric solution, evident in the large phase centre offset \citep[see Table~D.1 in][]{overview_arks} resulting from these data being collected prior to the release of accurate Gaia astrometric stellar solutions.

The observations of HD~15115, HD~32297, and HD~109573 have phase centre offsets from their Gaia stellar positions in both RA and Decl. smaller than 20~mas, showing that the star was positioned at (or very close to) the phase centre throughout their observations.
We present in Table~\ref{tab:checksources} the best-fit check source positions derived in both the visibility and image domain, i.e. with {\tt uvmodelfit} and {\tt imfit} respectively, as well as the best-fit disc centre offsets for HD~15115, HD~32297, and HD~109573.
The offset uncertainties associated with these three targets are very small, in most cases just a few milliarcseconds, whereas their disc centres are offset by many tens of milliarcseconds.
These three systems are amongst the nine ARKS targets with `check source' observations.
Check source observations enable the quantification of the typical astrometric uncertainties during ALMA observations, which we estimated for our science targets as described in Appendix~\ref{app:checkSources}.
The check source positions have tightly constrained locations consistent with no offset, and uncertainties that are smaller than the best-fit disc centre uncertainties.
This comparison suggests that HD~15115, HD~32297, and HD~109573 have disc centres significantly offset from their stellar locations that cannot be explained by instrumental/calibration errors.
We therefore concluded that these offsets are due to a physical offset from the centre of these three discs and their stars, and discuss this further in \S\ref{sec:results}.

\subsection{Assessment of asymmetric (sub)millimetre emission}
\label{sec:asymmetryassessments}
Appendix~\ref{sec:AppendixSources} presents the images analysed in this paper for emission asymmetries. 
We note here that for 23/24 sources, we use the standard ARKS continuum data, except for HD~107146, for which we use the Band~6 data alone.\footnote{The ARKS image for HD~107146 includes both Band~7 and Band~6 data imaged in {\tt tclean} with multi-frequency synthesis (mfs) and achieves exceptional S/N. 

However, due to an issue with the short baselines in the Band~7 data which led to significant fringes in the final cleaned image \citep[that could not be corrected for during the cleaning process; see][]{Imaz-Blanco2023} we removed the higher-frequency data and re-imaged the Band~6 data only. Whilst the Band~6 data have slightly worse resolution, the overall S/N is better than the Band~7 data and thus we find it more reliable for studying asymmetric emission.}
We primarily utilised two empirical methods to investigate whether systems host emission asymmetries. 
With these methods, we define significant asymmetries as those where we find evidence of residual asymmetric emission exceeding $5\sigma$, and tentative asymmetries where there is evidence of residual asymmetric emission in the range 3--5$\sigma$.

\subsubsection{Method one: Two-dimensional asymmetry analysis}
We conducted image self-subtraction analysis in the three disc symmetry-axes on the cleaned images of all 24 systems.
These disc symmetry axes are the major axis (which we use interchangeably here with position angle; PA), the minor axis ($90^\circ$ anti-clockwise to the PA), and the rotation axis (i.e. symmetry about a rotation of $180^\circ$).
Measuring for these non-axisymmetries requires knowledge of the image coordinate centre, which we fix to be the Gaia DR3 location of each system's star,\footnote{All ARKS systems (per observational execution block) had their data shifted to this same reference location, which was conducted in {\tt CASA} using the {\tt fixvis} and {\tt fixplanets} tasks; see \citet{overview_arks} Table~D.1 for further details.} unless these systems are host to a significant stellocentric offset (HD~15115, HD~32297 and HD~109573), in which case we shift the centre by this offset to hunt for asymmetries in their brightness distributions not caused by their offset alone, or the best-fit disc centres derived in \citet{overview_arks} provided significantly smoother residual maps in comparison to those produced assuming the image centre was the Gaia DR3 stellar location (HD~131488, and HD~131835), given the best-fit disc centre can be subtly shifted due to the presence of disc substructures; see \citet{ver_arks}.
The major and minor axis self-subtractions also require knowledge of the system position angles, and we use those provided by symmetric Gaussian model fits to each system's visibilities as presented in Table~B.1 of \citet{overview_arks}.
To produce these residual maps, we simply mirrored or rotated each image about the coordinate centre and subtracted the mirrored or rotated map from un-mirrored or un-rotated map.\footnote{For this analysis, we utilised the non-primary beam-corrected images for each source, but scale the contour levels by the primary beam correction, such that the contours reflect the underlying significance of residual features.}
Whilst this method is effective at finding strong asymmetries, it has the downside of inflating the noise properties of the residual images by $\sqrt{2}$. In addition, since these images are not spatially averaged, these properties can make subtler asymmetric features less pronounced.
Using this method, we only find strong evidence of asymmetric emission towards HD~121617 \citep[the most asymmetric disc in ARKS, as discussed in][]{hd121617_arks, line_arks, vortex_arks}, as well as tentative evidence in the cases of HD~9672, HD~10647, HD~39060, HD~92945, HD~107146, HD~131488, HD~131835, and HD~218396, each of which we discuss in \S\ref{sec:results}.
In each case, we verified that the observed asymmetric features are robust to whether we use the coordinate centres derived in \citet{overview_arks} as outlined above, or those derived in the ARKS axisymmetric parametric modelling papers, i.e. \citet{rad_arks} and \citet{ver_arks}.

\subsubsection{Method two: One-dimensional asymmetry analysis}
To search for subtler asymmetries we computed azimuthal, major axis, and minor axis profiles by averaging emission over particular regions of ARKS target images.
This process effectively integrates 2D emission into 1D profiles, which allowed us to detect fainter asymmetric emission features.
To produce these profiles, we de-projected the image data into system-specific bases, which requires knowledge of a target's bulk disc position angle, inclination, and reference coordinate centre. 
We adopted the values presented in Table~B.1 from ARKS~I \citep{overview_arks} for these parameters, and present these here for completeness in Table~\ref{tab:sourceproperties}.
We discuss each of these asymmetries per source in \S\ref{sec:results}.
In all cases, we considered residual emission found at levels above $5\sigma$ in intensity, or integrated (accounting for the number of independent resolution elements within the integrated region) to be conclusive evidence of the presence of an asymmetry. 
We considered residual emission, either in intensity or integrated, at the level $3-5\sigma$ to be tentative evidence of asymmetric emission.
We verified the robustness of these asymmetries to the reference coordinate centre by checking if these persist when we adopt the $\Delta$RA and $\Delta$Decl. values in the ARKS axisymmetric modelling papers, i.e. \citet{rad_arks} and \citet{ver_arks}.

Due to the diversity of disc geometries, although we computed these profiles for all systems, some are much poorer at providing diagnostics of disc emission asymmetries, and so we do not utilise these in our analysis.
For example, we do not interpret the azimuthal profiles of highly inclined or edge-on discs since their de-projections are too uncertain.
Details of the profiles used in the analysis of each system are provided in \S\ref{sec:results}.

A point that we highlight here is that the subtraction of submillimetre galaxies (SMGs) from the ARKS data pre-imaging \citep[as noted in ARKS~I][]{overview_arks} might remove some non-axisymmetric emission from each data set.
Five systems were shown to host SMGs that are spatially co-located within their discs: HD~76582, HD~95086, HD~107146, TYC~9340-437-1, and HD~218396 (HR~8799).
In the case of HD~107146, the subtracted emission originates from the inner cavity of the disc, inside the region where we assessed that there is asymmetric emission and thus, these subtractions do not affect our conclusions.
In the case of HD~218396, one such clump is coincident with one of the clump features that we identified as a tentative asymmetry, and hence we do not ascribe significance to this being a disc-related asymmetry.

\subsection{Assessment of asymmetric scattered light emission}
We checked for any signs of asymmetries in near-IR scattered light observations. 
We refer to ARKS~V \citep{scat_arks} for a summary of the data availability and reduction process. 
Here we focus on targets for which the discs have been detected with the VLT/SPHERE instrument (\citealp{Beuzit2019}), either in total intensity or in linear polarimetry. 
For polarimetric observations, we used the $Q_\phi$ image \citep{scat_arks}. 
For observations in total intensity, we first computed an estimate of the telescope's point spread function from the cube of images, using a principal component analysis (automatically selecting the number of principal components so that 95\% of the variance is taken into account); see also \citet{Soummer+2012}. 
We then subtracted this estimated point spread function to each frame of the cube, de-rotated them according to their parallactic angle at the time of the observations, and median-stacked the cube.
Afterwards, with the final image of the disc, we mirrored it along its minor axis \citep[with the position angle taken from Table~\ref{tab:sourceproperties}, see also ARKS~I]{overview_arks}, and subtracted it from the original image. 
If a target has both total intensity and linear polarimetric observations, both are used. The residual maps are shown in the top gallery of Fig.\,\ref{fig:selfsubgallery5}, where each map is convolved by a 2D Gaussian with a standard deviation of 1.5\,pixels to reduce the shot noise.

To further investigate the presence of asymmetries in these residual maps, we computed radial profiles, shown in the lower gallery of Fig.\,\ref{fig:selfsubgallery5}. 
The profiles are computed in wedges along the major axis of the discs (indicated on top of the residual images), and we computed the mean in radial bins, on both sides of the disc's minor axis. 
To estimate the uncertainties, we computed the standard deviation in the same radial bins on the uncertainties maps: we used the $U_\phi$ image for polarimetric observations, while for total intensity datasets, we used the same principal component analysis as mentioned before but de-rotate the cube in the opposite direction (to cancel out any disc signal).

\section{Results}
\label{sec:results}
\subsection{Asymmetries and offsets in ARKS: System-by-system analyses}

Using the methods outlined in Sect. \ref{sec:offsetsAsymms}, we identified asymmetries and offsets in ten ARKS targets (at levels above $5\sigma$), which we detail individually below. These are shown in Fig.~\ref{fig:gallery}.

\begin{figure}
    \centering
    \includegraphics[width=1.0\linewidth, clip, trim={0cm 0cm 0cm 0cm}]{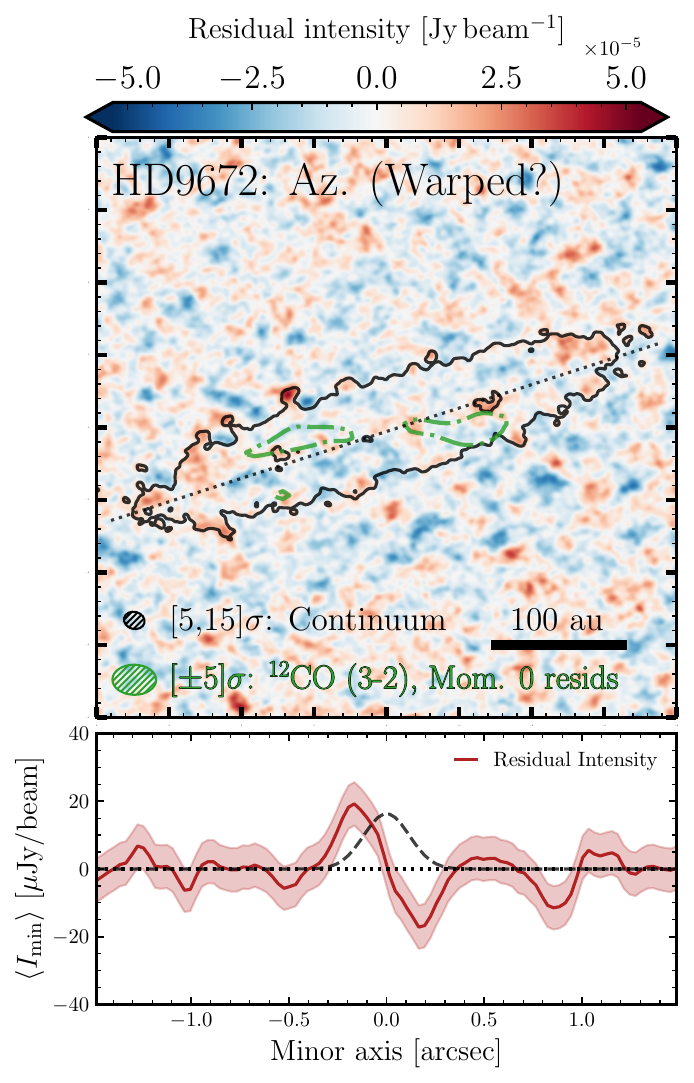}
    \caption{Top: HD~9672 residual map as presented in ARKS~III, \citet{ver_arks}, with the ARKS continuum image shown in solid black contours, and the $^{12}$CO (3-2) moment-zero residuals overplotted as thick dash-dot green contours, as presented in \citet{Hughes2017}, cropped to a $\pm4''$ region of the sky. The continuum disc position angle is shown as the black dotted line across the disc. Clean beams, contour scales, and a scale bar are shown on the map. North is up; east is left. Bottom: Minor axis residual intensity profile across a major axis averaged region about the ansae regions, shown in solid red (after reversing the direction of the western ansae). Errors are shown at the ${\pm}1\sigma$ level, scaled to account for the mean and difference profiles.}
    \label{fig:HD9672}
\end{figure}

\subsubsection{HD~9672 (49~Ceti): An azimuthally asymmetric disc}
HD~9672 hosts a unique azimuthal asymmetry.
With reference to Fig.~\ref{fig:gallery}, we note that the peaks in the disc ansae appear shifted (i.e. rotated or twisted) relative to the disc major axis, i.e. these are on opposing sides of the disc as can be seen by comparing the peak locations relative to the black dotted line which marks the disc position angle. 
To confirm this twisted feature asymmetry, we analysed both the ARKS image, and the best-fit model-subtracted residual map presented in ARKS~III \citep{ver_arks}, the latter of which we present in the top panel of  Fig.~\ref{fig:HD9672}.
We constructed and compared averaged intensity profiles parallel to the minor axis of the disc, through both disc ansae (averaged over a major axis extent of $0.75''$).
After reversing the direction of the profile through the western ansa and averaging this with the profile through the eastern ansa, we show in the residual intensity profile (lower panel of Fig.~\ref{fig:HD9672}) that about the major axis there is a significant enhancement and decrement feature, showing the twisted feature to be significant.
Indeed, by integrating the modulus of the residual intensity profile between ${\pm}0.5''$, we found that the twisted feature has a significance of $5.2\sigma$.
The twisted feature is somewhat evident in the self-subtraction maps for HD~9672 (see Fig.~\ref{fig:selfsubgallery1}, in particular the minor axis map); however, the residual intensity profile is necessary to determine its presence as significant.

This analysis demonstrates that there is a significant azimuthal asymmetry in HD~9672's continuum disc, the only source in the ARKS sample to present this type of twisted asymmetric morphology.
To highlight how this asymmetric feature manifests at other submillimetre data, we overplot on the top panel of Fig.~\ref{fig:HD9672} the $^{12}$CO (3-2) moment-zero residual map presented in \citet{Hughes2017}, where a similar twisted feature is present.\footnote{To overplot these data, we shifted the image centre of the CO data by the phase centre offset noted in \citet{Hughes2017}.}
Although these archival data have a resolution that is two times lower than the ARKS continuum data and also have lower sensitivity, the CO residuals appear stronger, and more significantly twisted from the disc position angle than the continuum dust.
Whilst there are multiple physical interpretations of this azimuthal structure, including the possibility that there is a warp, a physical twist, or a spiral, the origin of this feature is nevertheless difficult to determine given the available data, and without recourse to detailed modelling efforts, which are beyond the scope of this paper.
Spiral patterns and twists are yet to have been reported in the literature in a debris disc at (sub)millimetre wavelengths \citep[though these have been seen in scattered light data, for the young discs around HD~141569A and TWA~7; see e.g.][respectively]{Clampin+2003, Ren+2021}; however, there are now a number of reported warps, i.e. asymmetric vertical or inclination distributions, that have appeared in both scattered light and (sub)millimetre observations, for example towards HD~39060 \citep[$\beta$~Pic; see e.g.][]{Mouillet+1997, Augeraeu+2001, Golimowski06, Dent14} and HD~110058 \citep[see e.g.][]{Kasper+2015, Hales+2022, Stasevic+2023}.
In these two systems, their warps have been interpreted as resulting from secular interactions with misaligned planets \citep[see also][]{Crotts+24b}, including scenarios where the disc's gravity plays a significant role in shaping the warp structure \citep[e.g. via secular-inclination resonances;][]{Sefilian+2025}.
In such models, the discs exhibit a (leading) two-armed spiral pattern, which is most prominent beyond the warp \citep{Nesvold2015, Farhat+2023, Sefilian+2025}. 
That other systems have been found to be warped which share morphological similarity to HD~9672 may suggest a preference to interpret HD~9672's structure as being warped, and likewise driven by similar phenomena.
Nevertheless, any conclusion that this disc is warped, or due to some other type of structure, requires further work.

\begin{figure}
    \centering
    \includegraphics[width=1.0\linewidth, clip, trim={0cm 0cm 0cm 0cm}]{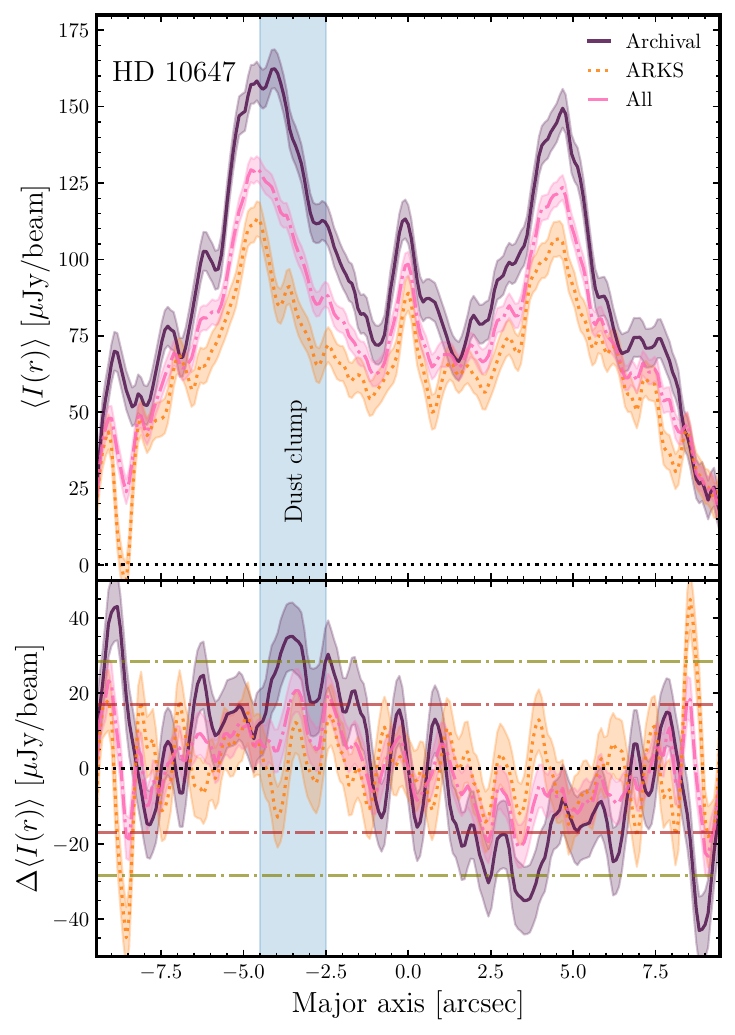}
    \caption{HD~10647 major axis profiles (top) for each of the archival-only, ARKS-only, or combined (`All') baselines, and their associated difference profiles (bottom) obtained by self-subtracting the profiles about $0.0''$. The major axis direction is negative towards the south-east of the disc. The red and green dash-dot lines  represent the ${\pm}3\sigma$ and ${\pm}5\sigma$ profile differences for the combined data;  the shading gives the error associated with each of the difference profiles, scaled by $\sqrt{2}$ to account for the self-subtraction. We also show  the ${\approx}40{-}80\,$au radial location (in blue) associated with q$^1$~Eri's dust clump from \citet{Lovell21c}.}
    \label{fig:HD10647_radial}
\end{figure}

\subsubsection{HD~10647 (q$^1$~Eri): A major axis asymmetry} \label{sec:q1eri}
The HD~10647 disc shows a significant major axis asymmetry, arising from an emission enhancement in its south-west ansa.
Although this asymmetry is visible in HD~10647's emission map, which hosts contours that show enhanced emission in the south-west ansa in comparison to the north-east (see Fig.~\ref{fig:gallery}), this feature is best characterised by its major axis profile that we present in Fig.~\ref{fig:HD10647_radial}.
In this figure we show the major axis intensity profiles (upper panel, averaged over the full minor axis of the disc) as a function of distance along the disc major axis from the central star for the ALMA data, separately for the archival data, ARKS-only data, or combined datasets and their self-subtracted intensity difference profiles (lower panel), and indicate the radial location of the reported clump \citep[from][]{Lovell21c} in blue.
Multiple independent peaks exceed $3\sigma$ along the south-east ansa region of the disc's combined data difference profile, and by integrating from 0 to either ${\pm}7.5''$, we found that the integrated difference profile is significant at the $5.5\sigma$ level. 
An interesting feature towards this source is that the significance of this asymmetry is strongly dependent on whether archival HD~10647 baselines are included within the imaged data.
For example, considering the archival-only (ARKS re-processed) data, the asymmetry significance increases to ${\approx}9.1\sigma$, with a net flux enhancement in the south-west of ${\approx}1.5\,$mJy, consistent with that reported in \citet{Lovell21c}.
However, this significance falls to ${<}2\sigma$ if only the newest (higher-resolution) ARKS-only data are considered.
Since the asymmetry is present in the fully combined data set, we report this here as significant.
The lack of a significant asymmetry in the ARKS-only data cannot be explained by the difference in observational resolution or sensitivity between the archival and ARKS-only data. We note these data achieved approximately the same sensitivity,  ($13.6\,\mu$Jy\,beam$^{-1}$ for the archival data, and $13.1\,\mu$Jy\,beam$^{-1}$ for the ARKS-only data), and by smoothing the ARKS-only data with a Gaussian kernel matching the restored (clean) beam properties of the archival data, we found no significant flux enhancement in the south-west.

Although we conclude that the major axis of HD~10647's disc is asymmetric, it appears less conclusive that this major axis asymmetry originates in an extended dust clump, as suggested by \citet{Lovell21c}.
The clump asymmetry interpretation of \citet{Lovell21c} showed broad agreement between the ALMA data and the HST images \citep[which shows the scattered light is extended asymmetrically further out in the north-west of the disc, presented also in][]{scat_arks}. 
The clump hypothesis suggested that the clump was produced as a Neptune-analogue planet migrated outwards and trapped planetesimals in its 2:1 mean motion resonance, similar to that expected in the Solar System \citep[see e.g.][]{Fernandez1984}.
Although this scenario remains consistent with the combined ARKS data, that the asymmetry significance has fallen suggests alternative hypotheses should be considered, in particular those that may explain the transience of this feature, for example, if this may be more plausibly due to planetesimal collisions \citep[see e.g.][]{Jones+2023}, or alternatively, something external to the disc, for example a variable background source (such as a variable star or a variable submillimetre galaxy).
Consequently, longer term monitoring of this feature is warranted to investigate its temporal behaviour.

\subsubsection{HD~15115: An eccentric disc}
HD~15115 is host to a stellocentric offset, as initially discussed in \S\ref{sec:offsets}.
From the self-subtraction map in Fig.~\ref{fig:selfsubgallery1} there is no evidence of emission enhancements in any of the three asymmetry axes considered.
Given the highly inclined geometry of this source, we produced a major axis profile that showed tentative signs of an emission asymmetry; however, this was not robust to the choice of coordinate centre (within the range implied by this source's RA and Decl. offsets; see Table~\ref{tab:sourceproperties}). 
We likewise found no evidence of an emission asymmetry about the minor axis. 
This means that, although the system hosts a significant offset, it presents as a symmetric disc in emission relative to the disc centre.

From the derived stellocentric offsets in Table~\ref{tab:checksources}, we found that the inferred pericentre direction is towards the western ansa.
However, due to the disc's highly inclined geometry, there is a large range of pericentre directions (i.e. argument of pericentre) and by extension, eccentricities, compatible with the data. 
Nevertheless, by assuming that the pericentre direction is aligned with the plane of the sky and oriented along the disc major axis, via the disc geometry the derived offsets imply a lower bound on the forced eccentricity of HD~15115 of $e_f>0.02$.
Although it is possible to place an upper bound on this value, this is subject to more uncertainty and will be constrained by future modelling work (Lovell et al., in prep.).

The significant offset between the disc centre and stellar location for this system means that, by definition, it is eccentric and adds to the small but increasing number of (sub)millimetre-resolved eccentric discs.
There is nevertheless a possible inconsistency between the ALMA-derived eccentricity and the scattered light emission. 
For example, although the ALMA offset analysis suggests that the pericentre direction is pointed westwards, the scattered light emission is dominated by asymmetric emission strongly extended westwards \citep[see e.g.][]{Kalas07, Schneider14} which may instead suggest the pericentre direction is directed eastwards, for example based on the models of \citet{Lee16}.
One plausible hypothesis may be that a recent collision in the disc occurred closer to orbital apocentre (rather than at orbital pericentre, where collision rates are typically more frequent), resulting in small grains being scattered preferentially westwards. 
Indeed, in \citet{Jones+2023} impact sites on both sides of the star were modelled to re-produce the scattered light morphology of the disc, which may result in the observed multi-wavelength morphology of HD~15115.
Our analysis of the SPHERE total intensity observations (see Fig.~\ref{fig:selfsubgallery5}) further confirm that the disc is asymmetric at near-IR wavelengths, with the west side brighter than the east side. 
Given the low signal-to-noise ratio of the SPHERE data, \citep{scat_arks} assumed a centro-symmetric model to reproduce the observations, which cannot be used to further investigate the location of the pericentre.
Further work is needed to address this potential difference between the submillimetre and scattered light analysis.

\subsubsection{HD~32297: An eccentric disc}
The debris disc of HD~32297 is host to a stellocentric offset, and we thus interpret this as being eccentric.
This finding is supported by observations of HD~32297 in ALMA Band~8 (Luppe et al., in prep.) in which the fitted RA and Decl. offsets match those in ALMA Band~7.
This system, similarly to HD~15115, also hosts no significant asymmetries in its emission along its major or minor axis.
The symmetry in these profiles is corroborated by HD~32297's self-subtraction maps in Fig.~\ref{fig:selfsubgallery1} which are measured about the disc centre. 
We note that if the stellar centre is instead adopted, strong self-subtraction residual emission artefacts are present given the significant stellocentric offset.
Nevertheless, there is a tentative minor axis asymmetry at the level of 3--4$\sigma$, that presents as an emission enhancement on the southern side of the disc, as noted in Table~\ref{tab:tentasymmetries} suggesting future observations may reveal there are subtle emission asymmetries in this disc.

From the derived stellocentric offsets in Table~\ref{tab:checksources}, we found that the inferred pericentre direction is towards the south-west ansa.
However, due to the disc's edge-on geometry, there is a large range of pericentre directions, and by extension, eccentricities, compatible with the data. 
Nevertheless, by assuming that the pericentre direction is aligned with the plane of the sky and oriented along the disc major axis, via this geometry the offsets imply a lower bound on the forced eccentricity of HD~32297 of $e_f>0.03$.

Our finding that HD~32297 is eccentric at submillimetre wavelengths is in good agreement with previous interpretations of scattered light observations.
One example is from the HST images of HD~32297 which shows large extension vertically northwards from the disc \citep{Schneider14}.
The HST data presents a strong match to the `double-wing' morphology as produced both by the models of \citet{Lee16}, resulting from perturbations from a highly eccentric internal planet, or by the models of \citet{Jones+2023} via a giant impact scenario. 
In addition, the SPHERE polarimetric data in scattered light presents a strong radial extension northwards, implying further that the pericentre direction is preferentially towards the south-west of the disc, i.e. in the same direction that we derived here based on the analysis of the stellocentric offset. 
This result is best observed in the polarimetric residual map of Fig.\,\ref{fig:selfsubgallery5} (top, third panel from the left); the disc is brighter in the south-west direction, close to the star, but the asymmetry is reversed farther out ($r \geq 0.8\arcsec$). 
On the north-east side, the disc is dimmer at short separation but becomes brighter beyond $r \geq 0.8 \arcsec$.
A similar pattern can be seen in the total intensity residual map, though there is strong residual speckle noise in the inner regions. 
Although \citet{scat_arks} modelled the disc assuming a circular birth ring, these SPHERE residuals are suggestive that HD~32297 hosts an eccentric disc, with a pericentre direction towards the south-west side.
\cite{Crotts24} and \cite{Duchene+2020} come to a similar conclusion based on GPI data, measuring modest eccentricities of 0.04 and 0.05, respectively, in an otherwise axisymmetric disc, which appear to be in very good agreement with our analysis of these ALMA data.

\subsubsection{HD~39060 ($\beta$~Pic): A major axis asymmetry}
\begin{figure}
    \centering
    \includegraphics[width=1.0\linewidth]{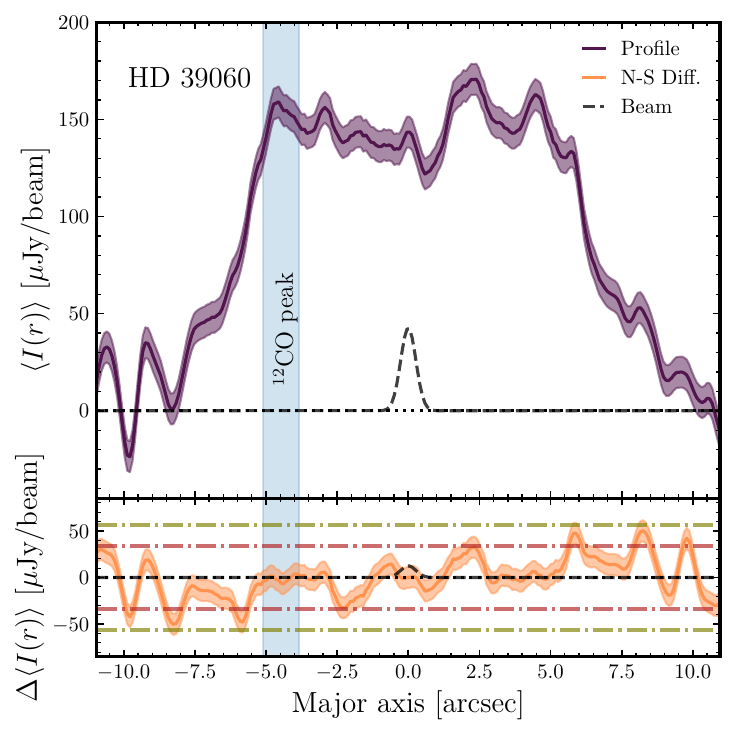}
    \caption{HD~39060 ($\beta$~Pic) major axis profile (top) from the south (minus major axis direction) to the north (positive radial direction), and its associated major axis difference profile about the image centre (bottom). 
     We show in red and green dash-dot lines ${\pm}3\sigma$ and ${\pm}5\sigma$ profile differences, and in amber the error associated with the difference profile, scaled by $\sqrt{2}$ to account for the uncertainty in the self-subtraction of the profile. We also show the ${\approx}85\,$au radial location (in blue) associated with $\beta$~Pic's known CO clump from \citet{Dent14}.}\label{fig:HD39060_radialvertical}
\end{figure}

The HD~39060 ($\beta$~Pic) disc is host to a significant asymmetry along its major axis.
This presents as a positive emission peak in the minor axis self-subtraction map in Fig.~\ref{fig:selfsubgallery2}, where this appears in the northern ansa of the disc.
By averaging emission over the disc minor axis ${\pm}2''$ from the mid-plane, and extracting the disc major axis intensity profile, as plotted in Fig.~\ref{fig:HD39060_radialvertical}, we found that emission is enhanced towards the north of the disc versus the south. 
We obtained a major axis difference profile for HD~39060 in the same manner described in \S\ref{sec:q1eri}.
By integrating the difference profile between of 0--10'' (accounting for the beam size) or equivalently, the modulus of the integral from -10--0'', we found this total flux enhancement at the level of $5.5\sigma$.
This northern enhancement is on the opposite side of the disc to the known CO, CI and dust clumps, and the `cat's tail' \citep[see e.g.][]{Dent2014, Cataldi2018, Han+2022, Rebollido+2024} which we discuss further below, which for the CO clump, we indicate in blue on Fig.~\ref{fig:HD39060_radialvertical}.
Separately, HD~39060 shows tentative evidence of a minor axis asymmetry, at the level of $3{-}4\sigma$, with the western side presenting brighter on the (northern) side of the disc from which the cat's tail protrudes \citep[also see][]{Rebollido+2024}.

The SPHERE residual map in Fig.~\ref{fig:selfsubgallery5} shows a unique asymmetric pattern for HD~39060, as we see positive and negative patterns on each side of the disc's minor axis. 
Since there is a known warped component in the disc and we used the position angle inferred from the submillimetre ARKS data, this suggests that the SPHERE polarimetric data are not probing the same component as the ALMA observations.
However, the scattered light and mid-infrared observations that are sensitive to HD~39060's extended emission may corroborate the asymmetry that we detect in these data.
For example, when imaged with HST, JWST and the University of Hawai'i Telescope (UHT), HD~39060 is always seen to be more extended towards northwards \citep[see][]{Kalas+1995, Golimowski06, Rebollido+2024}.
Indeed, in a giant impact scenario, the clumps in the south may trace a collisional impact location, and the northern extended emission may trace the radii out to which this material is preferentially ejected. 
Therefore, whilst the SPHERE data may be tracing a different asymmetry to ALMA, the northern extension traced by HST, JWST and UHT may be tracing the same physical feature.
Although we found no evidence for a significant stellocentric offset with the available ALMA data, this northward extension may imply that the HD~39060 debris disc is eccentric, which future observations should aim to constrain.
We note that in the theoretical work of  \citet{Smallwood2023}, which modelled the HD~39060 system including both known planets, it was shown that the disc may also be eccentric, resulting from interactions with these two planets, which may also drive the more famous warp asymmetry. 
Further work to study this disc's multiple asymmetries is thus warranted.

\subsubsection{HD~61005: A minor axis asymmetry}
\begin{figure}
    \centering    \includegraphics[width=1.0\linewidth]{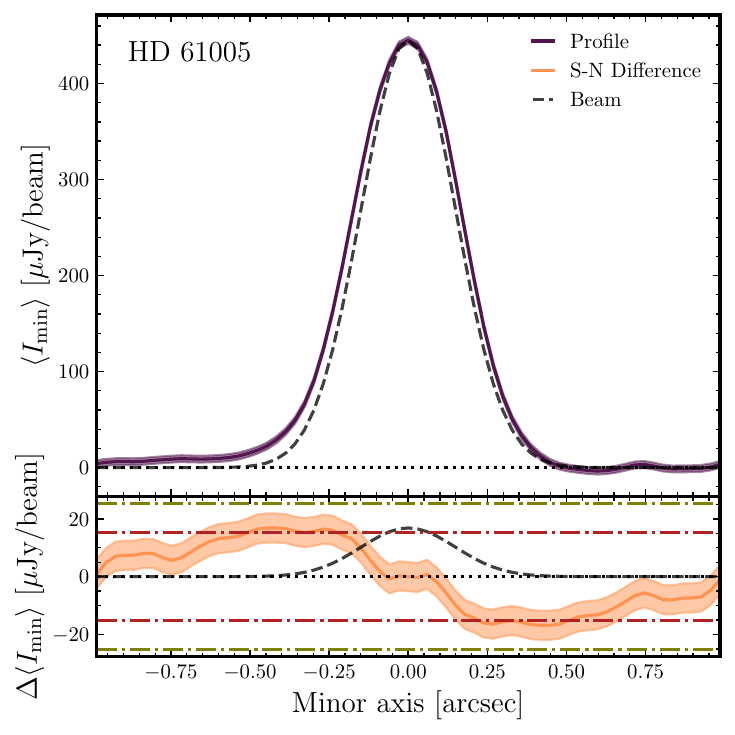}
    \caption{
    HD~61005 minor axis profile (top) from the south (minus minor axis direction) to the north (positive minor axis direction), and its associated difference profile about the image centre (bottom). 
    We show as red and green dash-dot lines the $\pm 3\sigma$ and $\pm 5\sigma$ profile differences, and in amber the error associated with the difference profile, scaled by ${\sqrt{2}}$ to account for the uncertainty in the self-subtraction of the profile.}
    \label{fig:HD61005_radialvertical}
\end{figure}

The HD~61005 disc is host to a minor axis asymmetry -- the only ARKS source to host such a feature at a significant level -- in which the southern half of the disc is enhanced in emission in comparison to the north.
We present the minor axis profile, produced over a major axis averaged extent of ${\pm}3.8''$, and the associated minor axis difference profile for HD~61005 in Fig.~\ref{fig:HD61005_radialvertical}, in the upper and lower panels respectively.
An enhancement exceeding $3\sigma$ in intensity is evident on the south side of the disc over a relatively broad span of ${\approx}2{-}3$ beams, which integrated, is $5.6\sigma$ significant.
By averaging over ${\pm}1''$ regions (parallel to the minor axis) along the disc major axis, in addition to the minor axis asymmetry, we found tentative evidence at the ${\approx}4\sigma$ level that the east side of the disc is brighter than the west.

HD~61005 is known as `The Moth' \citep{Hines+2007} as it is host to an iconic vertically extended (halo) emission distribution towards the south of the disc, as imaged in scattered light \citep{Schneider14, Olofsson+2016}, which appear like `swept-back wings'.
The asymmetry about this axis is clearly detected both in total intensity and linear polarimetry in the SPHERE observations, as seen in Fig.~\ref{fig:selfsubgallery5}.
In both datasets, the east side is brighter than the west side, and this holds true close to the minor axis, the major axis, and the swept-back wings. 
This suggests that the disc might be eccentric, with the pericentre located in the north-east side, possibly close to the semi-minor axis on the front side of the disc (e.g. \citealp{Olofsson+2016}, and consistent with the eccentric disc models of \citealt{Lee16}), though we note that no significant stellocentric offset was observed for this source with the ALMA data.
That we observe an asymmetry on this same side of the disc in thermal emission with ALMA suggests that the ARKS data may be sensitive to the extended halo emission as yet only seen at shorter wavelengths.
Comparatively, for both HD~32297 and HD~61005, \citet{MacGregor+2018b} reported the detection of an extended halo emission towards the edges of the major axis, previously only traced by scattered light.
As with \citet{MacGregor+2018b}, we remain agnostic as to whether the larger grains traced by the submillimetre ALMA data favour or disfavour the vertical asymmetry as being forced from ISM-gas ram pressure interactions with HD~61005's debris disc, without conducting further modelling of the disc, but we note that such modelling is now required to investigate this scenario. 

\subsubsection{HD~92945: An arced disc}
\begin{figure}
    \centering
    \includegraphics[width=1.0\linewidth, clip,trim={0cm 2cm 1.2cm 3.1cm}]{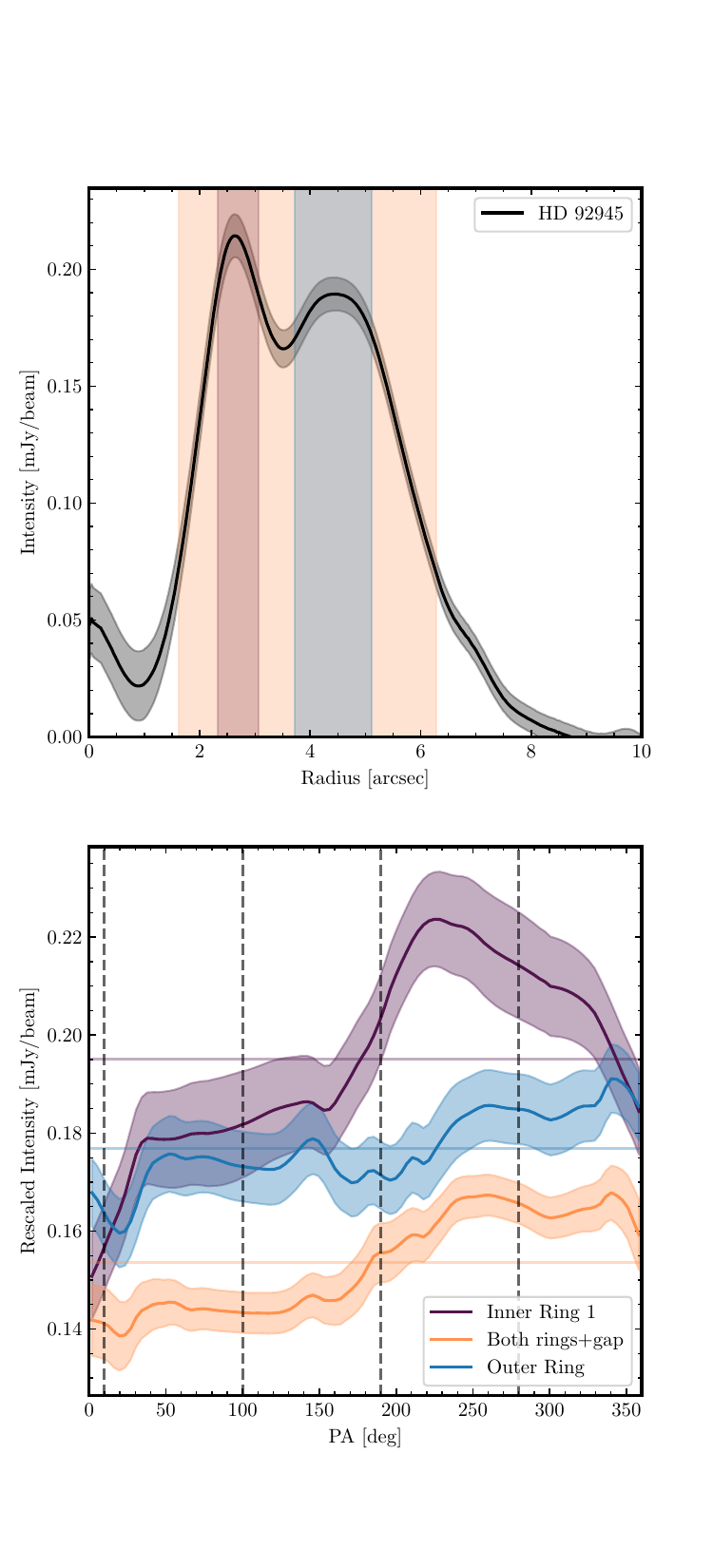}
    \caption{HD~92945 radial profile (top) and azimuthal profiles (bottom, as per indicated regions in top panel). The (top) shaded regions indicate the radial span over which profiles are extracted. The  shaded regions about each profile in the bottom panel show ${\pm}1\sigma$ error ranges. The vertical dashed lines indicate the PA, PA+90$^\circ$, PA+180$^\circ$, and PA+270$^\circ$  where the PA of HD~92945 is $100.0^\circ$ (see Table~\ref{tab:sourceproperties}), the horizontal purple, blue, and amber lines represent the mean profile for the inner ring, outer ring, and both ring profiles, respectively.}
    \label{fig:HD92945_radAzim}
\end{figure}

The HD~92945 disc is host to an arc asymmetry, that is strongest on its innermost ring, in the south-west of the disc.
We note that HD~92945 has been identified as host to two rings separated by a gap in between \citep{Marino19}, as corroborated by the ARKS data re-reduction \citep[see][]{overview_arks, rad_arks}.
HD~92945's arc asymmetry is amongst the strongest asymmetries we find in the sample, with this one clearly visible in its gallery image in Fig.~\ref{fig:gallery}, and plausibly too in its self-subtraction maps, i.e. Fig.~\ref{fig:selfsubgallery2}.
To quantify this brightness enhancement, we constructed azimuthal profiles around the disc over three radial spans, one for each of the two rings separately, and one spanning both rings and the gap.
These regions are shown in the radial profile of Fig.~\ref{fig:HD92945_radAzim} (top panel), alongside their corresponding azimuthal profiles (lower panel).
A strong enhancement is evident in the south-west of the disc in the inner ring (purple profile), which has a peak-enhancement at the level of $\approx30\%$ which spans a broad azimuthal extent of $\approx140^\circ$, which is resolved as an extended arc.
In comparison, the outer ring shows no evidence of a comparable enhancement, suggesting that the asymmetries in this disc are localised to, or dominated by, the arc in the inner ring.
We note that this asymmetry was first identified in \citet{Marino19}; however, since this could have been due to a background galaxy, this was not considered to be physically confirmed. 
However, recent scattered light observations with JWST NIRCam confirms that this asymmetry is also present in the distribution of small grains \citep{Lazzoni+25}, and hence here we report this asymmetry to be physically located in the disc.

There are multiple explanations to interpret the azimuthal asymmetry present in HD~92945's inner ring.
Given the large azimuthal extent of the arc, one possibility is that this is due to limb-brightening from the underlying disc being eccentric \citep[see e.g.][though we note the intensity enhancement might be too large for this to be feasible]{LynchLovell21, LovellLynch2023}.
An alternative, is that this is due to planet-disc interactions in the inner ring. 
\citet{Marino19} present models that include secular resonances that induce both the main gap and the asymmetry, and thus provide a plausible interpretation for this system.
Although not specific to the HD~92945 system, we note that the system TWA~7 has recently been identified to host similar sinusoid-like variations in its scattered light azimuthal profile \citep[see][]{Olofsson+2018, Ren+2021, Crotts+2025} due to direct planet-clearing, and thus an alternative explanation may be that this feature arises from a planet embedded in the disc.
In any case, deeper observations and further modelling of this target are warranted to investigate the nature of this arc, and its potential relationship with putative planets.

\begin{figure}
    \centering\includegraphics[width=1.0\linewidth, clip, trim={0cm 2cm 1.2cm 3.1cm}]{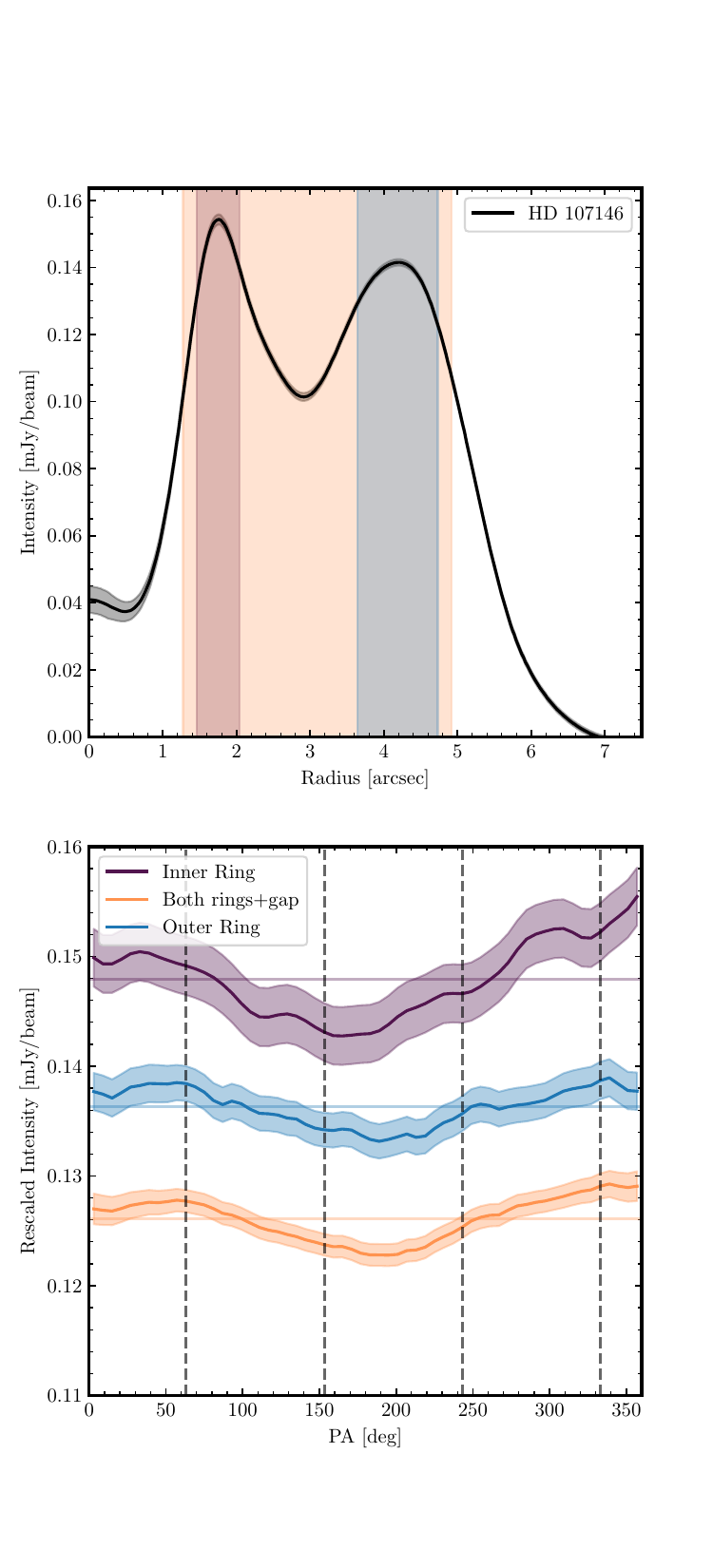}
    \caption{HD~107146 radial profile (top) and azimuthal profiles (bottom, as per indicated regions in top panel). The  shaded regions in the top panel indicate the radial span over which profiles are extracted. The  shaded regions about each profile in the bottom panel show ${\pm}1\sigma$ error ranges. The vertical dashed lines indicate the PA, PA+90$^\circ$, PA+180$^\circ$, and PA+270$^\circ$, where the PA of HD~107146 is $153.2^\circ$ (see Table~\ref{tab:sourceproperties}), the horizontal purple, blue, and amber lines represent the mean profile for the inner ring, outer ring, and both ring profiles, respectively.}
    \label{fig:HD107146_radAzim}
\end{figure}

\subsubsection{HD~107146: An arced disc}
The HD~107146 disc, like that of HD~92945, is host to an azimuthally extended arc asymmetry. 
In the HD~107146 disc, this is present in its northern side.
Another similarity with HD~92945, is that HD~107146 has also been identified as being host to rings and gaps \citep{Marino18, Imaz-Blanco2023}, likewise corroborated by the ARKS data re-reduction \citep[see ARKS I and II]{overview_arks, rad_arks}, although HD~107146 hosts three rings and two gaps (compared to HD~92945's two rings and single gap).
We recall that for this system, we have only used the available archival Band~6 ALMA data; see Sect. \ref{sec:asymmetryassessments}. 
In Fig.~\ref{fig:HD107146_radAzim} we show the radial profile for HD~107146 and three associated azimuthal profiles that span the innermost and outermost bright rings separately, and a third one that spans the full disc extent (i.e. also covering the gap between the rings, where this main gap is the combination of the two narrower gaps visible in higher-resolution data along with the third or centre ring).
These profiles demonstrate similar broad features as HD~92945, though in this case, there is evidence that the outer radial span contributes to the observed asymmetry, given the enhancement in the significance of the asymmetry in the profile which spans the largest radial extent across the disc (see the amber `Both rings+gap' profile). 
We measured a peak-trough enhancement of ${\sim}3\%$ for the full disc span, and a higher level of ${\sim}7\%$ in the inner disc when considered separately (though the latter of these has larger uncertainty by ${\approx}2\times$.
It is difficult to assess the extent of this arc, given that this feature appears to span almost fully around the disc (in an almost sinusoid-like pattern), and in this regard appears distinct to HD~92945, where the arc appeared more localised.
Neither SPHERE nor GPI scattered light data are available for HD~107146 to compare with this ALMA analysis, and we note that the HST study of HD~107146 \citep{Ertel+2011} presented no evidence of an arc asymmetry. 
The HST STIS scattered light image of HD~107146 is presented in \citet{scat_arks}.

The arc asymmetry in HD~107146's disc presents a number of interesting hypotheses as to its origin.
HD~107146 hosts a strong proper motion anomaly from a ${\sim}3M_{\rm Jup}$ body orbiting within 20\,au ($1.37''$) from its star \citep[see ARKS~I][]{overview_arks}, which lends weight to the possibility that the millimetre disc asymmetry may be tracing interactions with such a planet.
This means that one possibility is that this millimetre brightness modulation is due to the disc hosting an underlying eccentricity, driven by this inner planet \citep[since near face-on eccentric discs can produce sinusoidal-like emission variations in systems shaped by internal eccentric planets; see e.g.][]{LynchLovell21}. 
Although we found no evidence of a stellocentric offset in HD~107146's data, the ALMA configurations adopted to observe this system did not include long baseline configuration data (which were necessary to determine the stellocentric offsets in HD~15115, HD~32297 and HD~109573) and thus these data are intrinsically less sensitive to the presence of an offset.
An alternative possibility could be that a planet initially between the two brighter rings migrated inwards, creating the deeper gap between the rings, and trapped planetesimals in mean-motion resonances in the inner-ring \citep[as discussed by][]{Friebe+2022}, the latter of which may contribute to the observed azimuthal emission structure.
Finally, we highlight again the similarity to HD~92945, and thus the possibility here too of a planet embedded in the disc producing this azimuthal modulation \citep[via a similar comparison to][]{Crotts+2025}.
Further work, such as new modelling and follow-up observations, is now important to uncover which of these scenarios, or indeed another scenario, may best interpret the arc in HD~107146.

\subsubsection{HD~109573 (HR~4796): An eccentric disc}
The HD~109573 disc is the third system in ARKS that we determined to host a significant stellocentric offset, and thus, we interpret this as eccentric at submillimetre wavelengths. 
HD~109573 is geometrically distinct from HD~15115 and HD~32297; being less inclined, its minor axis is clearly resolved, meaning that its azimuthal structure can be more readily probed.  
By considering the azimuthal emission around the full radial extent of the ring, and separately on the disc's inner edge and outer edge, we found no evidence that the emission is significantly asymmetric about the disc centre, corroborated by Fig.~\ref{fig:selfsubgallery3} which shows HD~109573's self-subtraction residual maps to be solely from noise.
We note that if instead of the coordinates of the disc centre, the coordinates of the star are adopted, HD~109573's azimuthal profiles, and its self-subtraction maps, presented very strong asymmetries, supporting the conclusion that the disc is eccentric.

Whilst in the cases of HD~32297 and HD~15115 their geometries meant that deriving a pericentre direction and eccentricity was not possible (only lower bounds on $e_f$ could be derived), the resolved nature of HD~109573's disc means that a direct estimate of the pericentre direction and eccentricity can be made. 
Based on simple geometrical arguments, i.e. the direction vector from the disc centre to the stellar position (see the offset values in Table~\ref{tab:checksources}), we find that the pericentre direction is oriented approximately $17^\circ$ clockwise from north (conspicuously located at an azimuthal emission minima in the disc).
Assuming this pericentre location, based on the relative distance between the stellar position and the ring at pericentre and apocentre, we approximate a value of the disc eccentricity of $e_f\approx0.10$.
Whilst determining uncertainties on these values is beyond the scope of this study, a future ARKS paper (Lovell et al., in prep.) will model these data in detail, to extract stronger constraints on this disc's eccentric morphology.

The conclusion that HD~109573 is eccentric, and with a value of $e_f\approx0.10$, is largely corroborated by scattered light observations, which have measured eccentricities in the literature ranging from 0.01 to 0.08 \citep[see e.g.][]{Perrin+2015, Milli17,Olofsson2019,Milli+2019,Chen+2020,Crotts24}, although we note the value we derive here is on the upper end of this implied distribution.
In all such scattered light images, both the total intensity and linear polarimetry observations showed that HD~109573's disc surface brightness is asymmetric. 
The scattered light asymmetries are most evident in SPHERE polarimetric data that we present in the lower left panel of Fig.\,\ref{fig:selfsubgallery5}.
In the radial profiles for HD~109573, these are first positive, then negative along the north side, with the opposite pattern along the south side, a typical sign of an eccentric disc (as stated above, these same patterns appear in the submillimetre self-subtraction residuals, if the star's coordinates instead of the disc centre coordinates are adopted).
We also found that the pericentre direction implied by the disc centre offset towards the north (in the ALMA data) is consistent with modelling of HD~109573's scattered light emission \citep[see e.g.][]{Olofsson2019}.

From the disc's geometry, it is evident that the pericentre direction is not pointing towards one of the disc ansae;  it is perhaps less surprising that the disc ansae do not have strong brightness asymmetries \citep[as one might expect if the disc position angle was aligned with either the pericentre or apocentre direction of eccentric planetesimal orbits; see e.g.][]{Pan16, LynchLovell21}, and as pointed out in previous studies of HD~109573, for example \citet{Wyatt99} and \citet{Kennedy18}.
Nevertheless, minimal or absent emission enhancement in HD~109573's disc ansae may imply that HD~109573's eccentric disc has a different underlying density distribution of planetesimals to other eccentric rings which are host to strong ansae emission enhancements \citep[e.g. Fomalhaut, HD~53143, and HD~202628; see][]{MacGregor17, Faramaz19, Kennedy20, MacGregor2022}, as likewise postulated for this system by \citet{Kennedy18} with lower-resolution ALMA data of HD~109573.
Further modelling of this system is warranted to interpret these findings, and differences with respect to the wider eccentric debris disc population.

\subsubsection{HD~121617: An arced disc}
The HD~121617 disc hosts the most significant continuum asymmetry in the ARKS sample, evident in both its image; see Fig.~\ref{fig:gallery}, and self-subtraction maps; see Fig.~\ref{fig:selfsubgallery3}.
The self-subtraction maps show multiple ${>}3\sigma$ peak enhancements coincident with the arc in the south-west of the disc.
We do not present the azimuthal profile of HD~121617 here, since this is presented in ARKS~VIII \citep{hd121617_arks} Fig.2, in which the arc enhancement spans an angular extent of ${\sim}90^\circ$, and is brighter than the mean disc intensity by ${\sim}40\%$.
The arc itself also hosts an asymmetry, being brighter in its northern side, i.e. there is an azimuthal gradient in the intensity of the arc, as evident in the major-axis self-subtraction map (Fig.~\ref{fig:selfsubgallery3}). 
The SPHERE polarimetric observations of the disc show a pattern comparable to the one of HD~109573, with an inversion of the radial profiles (from positive to negative on the east side, and the opposite on the west side), indicative that this system has an eccentric ring in scattered light \citep[as reported at the level of $e=0.03$ by][]{Perrot+2023}, though we did not find evidence that this system was host to a stellocentric offset in the ALMA data, and so do not interpret this system as necessarily eccentric at submillimetre wavelengths.
We note that the brightness enhancements in the arc (detected by ALMA) do not present consistently in the near-IR scattered light observations.
We limit discussion to this source here since other ARKS papers are dedicated to further discussion of HD~121617 (see e.g. \citealt{line_arks},  \citealt{hd121617_arks}, and \citealt{vortex_arks}, which collectively model and analyse the structure of the continuum and gas kinematics in detail).

\subsection{Tentative asymmetries}
In three discs so far, we described the presence of additional tentative asymmetries, i.e. two in the minor axes of HD~32297, HD~39060, and one in the major axis of HD~61005.
In the wider ARKS sample we found evidence of tentative asymmetries in the discs of HD~84870, HD~131488, HD~131835, and HD~218396 that we briefly discuss here to motivate future observations.
We present the self-subtraction maps of these four systems in Fig.~\ref{fig:selfsubgallery55} \citep[alongside HD~95086 in Fig.~\ref{fig:selfsubgallery4} which we cannot confidently ascertain if this source is symmetric or asymmetric due to the imperfect-subtraction of a background source; see ARKS~I][]{overview_arks}.
Discussing these in turn, HD~84870 is host to a 3--4$\sigma$ clump in the south of the disc that appears in both the rotation subtraction and minor axis subtraction maps.
HD~131488 is host to residual emission features about its minor axis (see Fig.~\ref{fig:selfsubgallery55}), which may indicate the presence of a clump, or an emission enhancement in the western ansa.
HD~131835 is host to a ${\approx}3\sigma$ negative residual feature in its major axis and minor axis self-subtraction plots in Fig.~\ref{fig:selfsubgallery55} towards the south-east, near to its minor axis. 
Given the negative feature is co-located in both the major and minor axis self-subtraction maps, this may indicate the disc has a dearth of emission (an azimuthal gap) located in the inner disc \citep[in][this system is shown to have a two-component disc, with the first (inner) component brighter in thermal emission and fainter in scattered light, and the second (outer) component fainter in thermal emission, but brighter in scattered light]{rad_arks, ver_arks, scat_arks, hd131835_arks}.
HD~218396 (HR~8799) is host to two clumps in the north and north-west of the disc, that present most significantly in the major axis subtraction map.
Whilst the north-west clump is most likely due to the imperfect subtraction of a known background source \citep[see ARKS~I, our discussion in Sect. \ref{sec:asymmetryassessments}, as well as][where this bright point source is noted]{Faramaz2021}, the northern clump (present, just below 3$\sigma$) is not co-located with any known background sources.
Given the presence of multiple giant planets in the inner tens of au of HD~218396 \citep[see e.g.][]{Marois08, Apai+2016, Wertz+2017, Boccaletti+2024}, and the possibility that this clump may arise from planet-disc interactions with one of those, this clump in particular may warrant further investigation.

Overall, whilst we do not wish to overstate the significance of these tentative features (none of these exceed $5\sigma$), we report these given the possibility that future observations may confirm their presence.
For example, since some number of these present as unresolved clumps, these should be determined as to whether they are co-moving, to rule out the possibility that these are all due to background confusion \citep[see e.g. the method of][]{KennedyLovell2023}.

\section{Discussion}
\label{sec:discussion}

\begin{table}
    \caption{ARKS debris discs determined to be asymmetric.}
    \centering
    \begin{tabular}{l|c c }
    \hline
    \hline
      Name & Asym. Type & Class \\
     \hline
HD~9672 (49~Ceti) &Az.&Warp? \\
HD~10647 (q$^1$~Eri)&Maj.& Clump? \\
HD~15115&Off.& Eccentric  \\
HD~32297&Off.& Eccentric  \\
HD~39060 ($\beta$~Pic) &Maj.& ?  \\
HD~61005& Min.& ?  \\
HD~92945&Az.& Arc  \\
HD~107146&Az.& Arc   \\
HD~109573 (HR~4796)&Off.&Eccentric  \\
HD~121617&Az.& Arc \\
\hline
    \end{tabular}
    \label{tab:asymmetries}
        \tablefoot{Asymmetries with unknown or ambiguous classes are denoted with a `?'. All ten systems are presented as a gallery in Fig.~\ref{fig:gallery}.}

\end{table}

\subsection{Discussion of asymmetries and offsets in ARKS}
\subsubsection{Continuum asymmetries in debris discs are common}
Overall, we showed that 10/24 ($42\%$) ARKS targets are host to asymmetric emission or stellocentric offsets in their dust distributions, and that four further systems host tentative asymmetries.
We summarise these asymmetries and offsets in Table~\ref{tab:asymmetries} and Table~\ref{tab:tentasymmetries} noting in both their asymmetry classes, and in the first table, and where these are unambiguous, their physical class.
For each source, we have identified and briefly discussed plausible physical origins of the asymmetries, including planet-disc interactions, stellar-disc interactions (e.g. flybys), and collisions as plausible origins of those identified.

That nearly half of all ARKS systems host a significant continuum asymmetry suggests that asymmetries are common in the population of bright debris discs that the ARKS programme has analysed.
In ARKS~I \citep{overview_arks}, the sample's (cold dust) fractional luminosities ($L_{\rm cold}/L_\star$) are presented.
Excluding HD~95086 (for which we cannot determine the presence of a real asymmetry), we find that whilst 8/10 of the brightest discs, i.e. those with $L_{\rm cold}/L_\star>5\times10^{-4}$ host significant asymmetries, just 2/13 of the fainter discs with $L_{\rm cold}/L_\star<5\times10^{-4}$ host significant asymmetries.
We note that the two discs in the brighter population that do not host significant asymmetries are HD~131488 and HD~131835, which we showed tentative evidence that these may host asymmetries.
This bias implies that either the most luminous discs in the ARKS sample are intrinsically more asymmetric, or that with higher S/N data on the fainter discs, these discs may also host similarly asymmetric emission.
The latter of these possibilities is corroborated by scattered light observations, given the near-80\% fraction of debris discs measured as asymmetric in the GPI survey \citep{Crotts24}, i.e. it is plausible that debris discs are more likely to present as asymmetric than symmetric.
The former of these possibilities could  imply, however, that a number of asymmetries are driven by giant impacts, as collisional events not only produce asymmetric clumps and arcs, but overall raise the dust content of discs in general, and would thus be accompanied with a rise in $L_{\rm cold}/L_\star$ \citep[see e.g.][]{Jones+2023}.

Whilst debris disc asymmetries may be common, (sub)-millimetre wavelength debris disc asymmetries in general appear to be mostly very subtle \citep[this appears in contrast to scattered light asymmetries, which often appear far stronger;][]{Schneider14, Jones+2023, Crotts24}.
This may plausibly relate to the underlying physics and origins of asymmetries in debris discs, given the strong wavelength-dependence of their structures, the scales on which distinct asymmetries are observed, and the typical S/N required to observe asymmetric emission \citep[see further discussion in Sect. \ref{sec:intro}, and e.g. ][]{Matthews14, Hughes18}.
We note that in just one ARKS disc, HD~121617, was the self-subtraction map analysis sufficiently conclusive on the presence of significant asymmetry, whereas for all nine other asymmetries, we had to resort to additional analyses (though we note in the case of the offset systems, these host strong residual patterns if their stellar centres are used instead of their disc centres). 
This overall suggests that in general (sub)-millimetre debris discs can be understood as being dominated by axisymmetric morphologies, with a subset hosting fainter, non-axisymmetric features.
This means that modelling debris discs in the paradigm of axisymmetric emission generally remain a well-justified approach for interpreting their bulk properties \citep[][]{Kennedy2025, rad_arks, ver_arks}.

The subtlety of debris disc asymmetries appears in contrast to those observed in large, structured protoplanetary discs, as evident in the uniform analyses of ALMA data in, for example, the DSHARP survey \citep{Andrews2021} and the exoALMA survey \citep{Curone+2025}.
In these works, it is evident that in a number of cases (e.g. MWC~758, HD~135344B), the emission structures of some protoplanetary discs are near-dominated by asymmetric features (e.g. some of these sources have non-axisymmetry index values above 0.4), as would not be the case in any single ARKS debris disc, as evident in Figs.~\ref{fig:selfsubgallery1}, ~\ref{fig:selfsubgallery2}, and ~\ref{fig:selfsubgallery3}.\footnote{In the cases of HD~15115, HD~32297, and HD~109573, if these were instead considered about their stellar locations, then they would be considered hosts to strong non-axisymmetric residuals. The comparison we make is therefore only valid for disc-centred residuals, and thus should be accounted for in any comparison of the relative non-axisymmetries of eccentric discs.}
These differences between protoplanetary disc asymmetries and debris disc asymmetries may therefore highlight the importance of dust--gas interactions in driving the majority of strong asymmetries in circumstellar discs.
This appears to be corroborated by our finding that the system with the strongest continuum asymmetry (HD~121617) is also the most CO-rich disc in the sample \citep{gas_arks}, as well as our discussion of the possible bias towards CO-detections in asymmetric discs; see Sect. \ref{sec:disc_COasyms}.

\begin{table}
    \caption{ARKS debris discs determined to be tentatively asymmetric.}
    \centering
    \begin{tabular}{l|c}
    \hline
    \hline
      Name & Asym. Type \\
     \hline
     HD~32297 & Min \\
     HD~39060 ($\beta$~Pic) & Min \\
     HD~61005 & Maj \\
     HD~84870&Az and Min \\
     HD~131488&Maj  \\
     HD~131835&Maj and Min \\
     HD~218396 (HR8799)&Az  \\
    \hline
    \end{tabular}
    \label{tab:tentasymmetries}
    \tablefoot{Overlap with systems in Table~\ref{tab:asymmetries} demonstrate that some systems show tentative hints of asymmetries in multiple features.}
\end{table}

\subsubsection{A diverse suite of asymmetries in debris discs}
We identified four broad asymmetry types present in the data; four discs are azimuthally asymmetric, three discs have stellocentric offsets, two discs have major axis asymmetries, and one disc has a minor axis asymmetry. 
Whilst this is the largest sample of debris discs host to (sub)-millimetre asymmetries to date, these fractional subpopulations of asymmetries are too small to conclude that the presence of any one type of asymmetry is more common than any other.
This conclusion remains true even if the tentative asymmetry classes are included.

Importantly, the physical origins of some asymmetry types can be confused by their observing geometries.
For example, an arc or a clump in a highly inclined debris disc may present a major axis asymmetry observationally. 
This places a limitation on our ability to identify the physical origin of some asymmetries, and in particular, systems identified with either major or minor axis flux asymmetries.
Moreover, some physical asymmetries can manifest in degenerate observational signatures.
For example, as discussed earlier in the case of HD~9672, we demonstrated the presence of a significant azimuthal asymmetry, though we are unable to conclude on this being a warp, twist or spiral based on the available data.
This further challenges our ability to infer the frequency of any specific physical classes of debris disc asymmetries.

Nevertheless, in six cases we are more confident about the origins of their asymmetries.
For example, in the case of HD~92945, HD~107146 and HD~121617 these are spatially resolved as arc asymmetries, where each of these systems is host to a significant dust column enhancement over an azimuthally extended region of the disc.
Further, in the case of HD~15115, HD~32297 and HD~109573, the systems with offsets, such offsets can only be induced if these systems are eccentric.
The discovery of three eccentric systems at submillimetre wavelengths with ARKS observations doubles the number of previously resolved eccentric debris rings at these wavelengths.
We thus add HD~15115, HD~32297 and HD~109573 to the ranks of Fomalhaut \citep[HD~216956; see e.g.][]{MacGregor17, Matra17b, Kennedy20, Chittidi+2025, Lovell+25_Fom}, HD~202628 \citep[e.g.][]{Faramaz19, Kennedy20}, and HD~53143 \citep[e.g.][]{MacGregor2022, LovellLynch2023}.
A key distinction between these three eccentric discs in ARKS and the three previously known is the lack of any observed stellar emission towards HD~15115, HD~32297, and HD~109573, whereas stellar (sub)-millimetre emission is strongly detected towards Fomalhaut, HD~53143 and HD~202628.
This study therefore demonstrates the potential for identifying new eccentric discs with high-resolution ALMA data, even in the absence of detectable (sub)-millimetre stellar emission.

\textit{A lack of spiral patterns in debris discs?}
It is likely that even with the sensitivity and resolution of ARKS data, spiral patterns would have gone undetected. 
Although the asymmetry in HD~9672 disc's could be due to a spiral pattern, we see no strong evidence of this type of asymmetry in the ARKS sample of debris discs, nor in the (less sensitive) REASONS survey of debris discs \citep{MatraReasons25}.
This type of asymmetry therefore appears to be especially rare in debris discs.
This is in contrast to protoplanetary discs where multiple spiral morphologies have been detected at submillimetre wavelengths \citep[see e.g.][]{Andrews18, Huang+18b, Andrews20}.
Spiral patterns in protoplanetary discs are understood to be produced by hydrodynamical processes, i.e. gas-driven, which are then reflected in the dust \citep[see e.g.][]{Goldreich+1979, Bae+2018}. 
In debris discs, which are in general much more gas-poor than protoplanetary discs, the mechanisms driving spiral substructures are likely to be fundamentally different, for example resulting from planet-induced secular differential precession of planetesimal orbits \citep[see e.g.][]{Wyatt05, SendeLohne19, Farhat+2023}, in which case these would be transient in nature, or resulting from density waves launched at secular resonances (if any) in discs with non-zero mass, in which case spirals may be long-lived \citep{WardHahn1998, Hahn2003, Sefilian2022}.
Such features, however, may be damped due to viscosity effects and/or collisional activity \citep[see e.g.][]{Hahn2003, Nesvold2015}, making them less discernible in observations. 

\subsubsection{Are CO-rich debris discs more commonly asymmetric?}
\label{sec:disc_COasyms}
In the ARKS sample there are 6 CO-rich debris discs \citep[namely, HD~9672, HD~32297, HD~39060, HD~121617, HD~131488 and HD~131835;][]{gas_arks}, and 18 CO-poor debris discs (where no CO detections have yet been reported).
We found that the presence of CO gas in a debris disc may be a good indicator of whether its dust continuum will be asymmetric.
For example, whilst 4/6 of these CO-rich debris discs host significant continuum asymmetries, only 6/18 of the debris discs without a CO detection are asymmetric.
If the tentative asymmetries that we report are determined to be real, then these figures both rise to 6/6 and 8/18 respectively.
Whilst these numbers remain too small for us to conclude that this trend is significant, it remains indicative that where CO gas has been detected, continuum asymmetries are more likely to be found.
Plausibly, this may mean that hydrodynamic interactions between the gas and dust are driving at least some of these asymmetric features.
This possibility is further discussed in \citet{vortex_arks} where the arc in HD~121617 is modelled as the result of dust trapped in a vortex.

\subsection{Comparison with ARKS visibility modelling}
In this work, we have uniformly analysed ARKS images to assess the presence of asymmetries, almost solely analysing the ARKS continuum images.
In ARKS~II and ARKS~III \citep{rad_arks, ver_arks} all ARKS targets were fitted with parametric axisymmetric disc models in the visibility domain (ARKS~III only modelled the $i>66^{\circ}$ systems), producing best-fit models and residual maps.
Given the systematic difference in these two methods to analyse the ARKS sample, here we compare the conclusions drawn between our respective methods. 

For the three eccentric discs, HD~15115, HD~32297, and HD~109573, which we found to be axisymmetric but significantly offset, the parametric modelling of ARKS~III found consistent offsets were needed to fit these data, and thus our studies agree on this conclusion. 
We note that for HD~109573, the parametric modelling presented in ARKS~III finds residual emission excesses in the disc ansae that we did not find strong evidence of in this study.
We estimate that HD~109573 is highly eccentric ($e_f\approx0.10$), and thus will be host to a non-axisymmetric surface density distribution.  Therefore, by fitting and subtracting an axisymmetric ring model, i.e. with a different surface density distribution to that of an eccentric ring, these residual artefacts are likely to appear.
HD~109573 will be studied in a future ARKS paper to understand if these excesses can be accounted for with such an eccentric disc model (Lovell et al., in prep.).

For the three systems with arc asymmetries, we can only directly compare with the parametric modelling efforts in two cases i.e. HD~121617 and HD~92945, since we only used archival Band~6 data to analyse HD~107146, whereas ARKS~II analysed the Band~6 and Band~7 data combined. 
In \citet{rad_arks}, both HD~92945 and HD~121617 are host to significant residual emission co-located with the arc features that we identified in this work, and we thus find very good agreement between our analysis here and the parametric modelling.

For the four systems analysed via profiles for asymmetries, HD~9672, HD~10647, HD~39060, and HD~61005, we likewise find good agreement between our conclusions and those of ARKS~III \citep{ver_arks}.
For HD~9672, since we analysed the image domain of the best-fit parametric model residuals, our results agree with ARKS~III by definition.
For HD~10647, ARKS~III present residuals that have a number of $3\sigma$ peaks along the south-west inner edge, in the same locations that we integrated over to produce the significant asymmetry that we present here.
For HD~39060, the parametric models presented in ARKS~III find a number of residual peaks in the north of the disc, which also agree with the major axis asymmetry that we present here.
Although few $3\sigma$ peaks appear in the ARKS~III residuals for HD~61005, we note that an analysis of the residual map reveals, on the northern side of the disc, that there is a relative model oversubtraction (leading to an integrated negative flux) in comparison to the south where there is a relative undersubtraction (leading to an integrated positive flux).
We thus also find general agreement between the ARKS~III residuals and our conclusion that HD~61005 has an asymmetry along the minor axis.

Overall therefore, the findings of the asymmetries in this work are in very good general agreement with the results presented by the ARKS parametric modelling in ARKS~II and III.

\section{Conclusions}
\label{sec:conclusions}
We present the sixth paper of The ALMA survey to Resolve exoKuiper belt Substructures (ARKS).
ARKS is studying a sample of 24 debris discs at unprecedented resolution to understand their radial and vertical structures, including asymmetries.
In this work we  presented our study of the dust continuum asymmetries and offsets in ARKS data.
From our analyses we conclude the following:
\begin{enumerate}
    \item In the ARKS sample, 10 out of 24 systems are hosts to either a continuum asymmetry or an offset, with tentative signs of asymmetries in four further systems. Of the ten significant asymmetries, three host stellocentric offsets (non-zero eccentricities), three host arc asymmetries, two host major axis asymmetries, one hosts a minor axis asymmetry, and one hosts an azimuthal asymmetry with a twisted pattern. 
    Overall, these results demonstrate that there is a broad diversity of asymmetries in the ARKS sample.
    \item Our investigation suggests that continuum asymmetries are a common feature in debris discs, when observed with sufficiently high resolution and sensitivity, which prior to ARKS was inconclusive.
    Nevertheless, these (sub)millimetre asymmetries mostly present as subtle features. 
    We find that the presence of an asymmetry is correlated with the cold dust fractional luminosity in the sample.
    We also find a tentative enhancement in the fraction of systems hosting a continuum asymmetry and in those that are CO-rich.
    \item We  discussed plausible interpretations for the origins of these asymmetries, including planet--disc interactions, stellar--disc interactions, and collisions. We  compared the ARKS asymmetries to those present in protoplanetary discs, which in general appear to be both qualitatively and quantitatively distinct. 
    We highlighted upcoming planned work that will investigate systems in detail, and proposed future studies to follow up targets with new observations, modelling, and simulations to better constrain these asymmetric features and the processes driving them.
\end{enumerate}

\begin{acknowledgements}
We thank the anonymous referee for their careful review of our study, and many helpful suggestions which helped improved this manuscript.
JBL acknowledges the Smithsonian Institute for funding via a Submillimeter Array (SMA) Fellowship, and the North American ALMA Science Center (NAASC) for funding via an ALMA Ambassadorship. 
SM acknowledges funding by the Royal Society through a Royal Society University Research Fellowship (URF-R1-221669) and the European Union through the FEED ERC project (grant number 101162711). 
AMH acknowledges support from the National Science Foundation under Grant No. AST-2307920. 
EM acknowledges support from the NASA CT Space Grant. 
TDP is supported by a UKRI Stephen Hawking Fellowship and a Warwick Prize Fellowship, the latter made possible by a generous philanthropic donation. 
A.A.S. is supported by the Heising-Simons Foundation through a 51 Pegasi b Fellowship. 
Support for BZ was provided by The Brinson Foundation. 
MB acknowledges funding from the Agence Nationale de la Recherche through the DDISK project (grant No. ANR-21-CE31-0015). 
AB acknowledges research support by the Irish Research Council under grant GOIPG/2022/1895. 
CdB acknowledges support from the Spanish Ministerio de Ciencia, Innovaci\'on y Universidades (MICIU) and the European Regional Development Fund (ERDF) under reference PID2023-153342NB-I00/10.13039/501100011033, from the Beatriz Galindo Senior Fellowship BG22/00166 funded by the MICIU, and the support from the Universidad de La Laguna (ULL) and the Consejer\'ia de Econom\'ia, Conocimiento y Empleo of the Gobierno de Canarias. 
EC acknowledges support from NASA STScI grant HST-AR-16608.001-A and the Simons Foundation. 
This material is based upon work supported by the National Science Foundation Graduate Research Fellowship under Grant No. DGE 2140743. 
JPM acknowledges research support by the National Science and Technology Council of Taiwan under grant NSTC 112-2112-M-001-032-MY3. 
SMM acknowledges funding by the European Union through the E-BEANS ERC project (grant number 100117693), and by the Irish research Council (IRC) under grant number IRCLA- 2022-3788. Views and opinions expressed are however those of the author(s) only and do not necessarily reflect those of the European Union or the European Research Council Executive Agency. Neither the European Union nor the granting authority can be held responsible for them. 
JM acknowledges funding from the Agence Nationale de la Recherche through the DDISK project (grant No. ANR-21-CE31-0015) and from the PNP (French National Planetology Program) through the EPOPEE project. 
S.E. is supported by the National Aeronautics and Space Administration through the Exoplanet Research Program (Grant No. 80NSSC23K0288, PI: Faramaz). 
MRJ acknowledges support from the European Union's Horizon Europe Programme under the Marie Sklodowska-Curie grant agreement no. 101064124 and funding provided by the Institute of Physics Belgrade, through the grant by the Ministry of Science, Technological Development, and Innovations of the Republic of Serbia. 
This work was also supported by the NKFIH NKKP grant ADVANCED 149943 and the NKFIH excellence grant TKP2021-NKTA-64. Project no.149943 has been implemented with the support provided by the Ministry of Culture and Innovation of Hungary from the National Research, Development and Innovation Fund, financed under the NKKP ADVANCED funding scheme. 
LM acknowledges funding by the European Union through the E-BEANS ERC project (grant number 100117693), and by the Irish research Council (IRC) under grant number IRCLA- 2022-3788. Views and opinions expressed are however those of the author(s) only and do not necessarily reflect those of the European Union or the European Research Council Executive Agency. Neither the European Union nor the granting authority can be held responsible for them. 
SP acknowledges support from FONDECYT Regular 1231663 and ANID -- Millennium Science Initiative Program -- Center Code NCN2024\_001.
PW acknowledges support from FONDECYT grant 3220399 and ANID -- Millennium Science Initiative Program -- Center Code NCN2024\_001.

This paper makes use of the following ALMA data: ADS/JAO.ALMA\# 2022.1.00338.L, 2012.1.00142.S, 2012.1.00198.S, 2015.1.01260.S, 2016.1.00104.S, 2016.1.00195.S, 2016.1.00907.S, 2017.1.00167.S, 2017.1.00825.S, 2018.1.01222.S and 2019.1.00189.S.
ALMA is a partnership of ESO (representing its member states), NSF (USA) and NINS (Japan), together with NRC (Canada), MOST and ASIAA 
(Taiwan), and KASI (Republic of Korea), in cooperation with the Republic of Chile.
The Joint ALMA Observatory is operated by ESO, AUI/NRAO and NAOJ. 
The National Radio Astronomy Observatory is a facility of the National Science Foundation operated under cooperative agreement by Associated Universities, Inc. 
This research used the Canadian Advanced Network For Astronomy Research (CANFAR) operated in partnership by the Canadian Astronomy Data Centre and The Digital Research Alliance of Canada with support from the National Research Council of Canada the Canadian Space Agency, CANARIE and the Canadian Foundation for Innovation. 
SPHERE is an instrument designed and built by a consortium consisting of IPAG (Grenoble, France), MPIA (Heidelberg, Germany), LAM (Marseille, France), LESIA (Paris, France), Laboratoire Lagrange (Nice, France), INAF–Osservatorio di Padova (Italy), Observatoire de Gen\`eve (Switzerland), ETH Zurich (Switzerland), NOVA (Netherlands), ONERA (France) and ASTRON (Netherlands) in collaboration with ESO. SPHERE was funded by ESO, with additional contributions from CNRS (France), MPIA (Germany), INAF (Italy), FINES (Switzerland) and NOVA (Netherlands). 
SPHERE also received funding from the European Commission Sixth and Seventh Framework Programmes as part of the Optical Infrared Coordination Network for Astronomy (OPTICON) under grant number RII3-Ct-2004-001566 for FP6 (2004–2008), grant number 226604 for FP7 (2009–2012) and grant number 312430 for FP7 (2013–2016). 
The SPHERE data presented here is based on observations collected at the European Southern Observatory under ESO programme(s) 095.C-0298(A), 0101.C-0420(A), 598.C-0359(F), 098.C-0686(B), 096.C-0388(A), 0102.C-0916(B), 095.C-0273(A), 0101.C-0753(B), 0104.C-0436(B), and 098.C-0686(A, B).
We also acknowledge financial support from the Programme National de Plan\'etologie (PNP) and the Programme National de Physique Stellaire (PNPS) of CNRS-INSU in France. This work has also been supported by a grant from the French Labex OSUG@2020 (Investissements d'avenir – ANR10 LABX56). 
\end{acknowledgements}

\section*{Data availability}
The ARKS data used in this paper can be found in the \href{https://dataverse.harvard.edu/dataverse/arkslp}{ARKS dataverse}. For more information, visit \href{https://arkslp.org}{arkslp.org}.
Additional non-standard ARKS products used within this work are stored in \citet{DVN/CNMP7W_2025}, where we have also made available the code used to produce the self-subtraction plots presented in Figs.~\ref{fig:selfsubgallery1}--\ref{fig:selfsubgallery55}. This code is also stored on \href{https://github.com/astroJLovell/imageAsymmetryAnalysis}{github}.

\bibliographystyle{aa}
\bibliography{bibfile}

\begin{appendix} 
\section{Source data}
\label{sec:AppendixSources}
The ARKS data have been consistently imaged with a range of robust parameters (0.5, 2.0) and with uv-tapers for some low surface brightness, extended sources.
Asymmetric features can be more or less pronounced in different images, and thus   we catalogue which images were used to produce various maps and/or profiles in this study in Table~\ref{tab:imageanalyses}, and for each source their respective inclination, position angle (PA), and right ascension (RA) and declination (Decl.) offset values in Table~\ref{tab:sourceproperties}.

\begin{table*}
    \centering
    \caption{Images and image properties used to analyse the ARKS sample in this paper.}
    \begin{tabular}{l|ccc|ccc|ccc}
    \hline
    \hline
    Science Target && Gallery &&& Self-sub. &&& Profile &\\
    & R & UVT & $\sigma_{\rm RMS}$ & R & UVT & $\sigma_{\rm RMS}$ & R & UVT & $\sigma_{\rm RMS}$ \\
    & &[$''$] & [$\mu$Jy\,beam$^{-1}$] & & [$''$] & [$\mu$Jy\,beam$^{-1}$] & & [$''$] & [$\mu$Jy\,beam$^{-1}$] \\
    \hline
    HD~9672   & 2.0&-- &11& 2.0&-- &11& 2.0&--&11 \\
    HD~10647 (All)  & 2.0&-- &9.4& 2.0&-- &9.4& 2.0&--&9.4 \\
    HD~10647 (Archival) & 2.0&-- &13.6& --&-- &--& --&--&-- \\
    HD~10647 (ARKS) & 2.0&-- &13.1& --&-- &--& --&--&-- \\
    HD~15115  & 0.5&-- &6.5& 0.5&-- &6.5& --&--&-- \\
    HD~32297  & 2.0&-- &15& 2.0&-- &15& --&--&--\\
    HD~39060  & 2.0&$0.5$ &20&  2.0&$0.5$ &20& 2.0&$0.5$&20 \\
    HD~61005  & 2.0&-- &19& 2.0&-- &19& 2.0&--&19 \\
    HD~92945  & 2.0&$0.7$ &21& 2.0&$0.7$ &21& 2.0&$0.7$&21 \\
    HD~107146 & 2.0&-- &5.6& 2.0&-- &5.6& 2.0&--&5.6 \\
    HD~109573 & 0.5&-- &20& 0.5&-- &20& --&--&-- \\
    HD~121617 & 0.5&-- &11& 0.5&-- &11&--&--&-- \\
    \hline
    HD~84870  &--&--&--& 2.0&-- &12& --&--&-- \\
    HD~95086  &--&--&--& 2.0&$1.0$ &28& --&--&--\\
    HD~131488 & --&-- &--& 2.0&0.1 &11& --&--&-- \\
    HD~131835 & --&-- &--& 0.5&-- &11& --&--&-- \\
    HD~218396 &--&--&--& 2.0&$0.8$ &11& --&--&-- \\
    \hline
    \end{tabular}
    \label{tab:imageanalyses}
    \tablefoot{`R' here refers to the robust parameter of the image, and `UVT' to the angular size of the uv-taper. 14/15 images are standard ARKS products, all except HD~107146, which we detail in Sect. \ref{sec:results}. No profile analyses were used to quantify asymmetries towards HD~15115, HD~32297, HD~109573, and HD~121617. The lower half of the table presents target information for the systems with tentative asymmetries, including HD~95086, whose asymmetry is strong, but may not be physical. $\sigma_{\rm RMS}$ defines the RMS associated with each image (all values are presented to two significant figures).}
\end{table*}

\begin{table*}
    \centering
    \caption{Geometric and offset parameters used to analyse the ARKS data in this study.}
    \begin{tabular}{l|cccc}
    \hline
    \hline
    Science Target & Inc. & PA & $\Delta$RA & $\Delta$Decl. \\
    & [$^\circ$] & [$^\circ$] & [mas] & [mas] \\
    \hline
    HD~9672   &$78.7\pm0.2$ & $107.9\pm0.2$&$68\pm76$ & $-33\pm30$ \\
    HD~10647  &$77.8\pm0.1$ & $57.3\pm0.1$&$90\pm56$ & $28\pm51$ \\
    HD~15115  &$86.7\pm0.0$ & $98.5\pm0.0$ &$44\pm12$ & $-8\pm5$\\
    HD~32297  &$88.3\pm0.0$ & $47.5\pm0.0$&$19\pm4$ & $18\pm4$\\
    HD~39060  &$86.4\pm0.2$ & $29.9\pm0.1$&$-54\pm14$ & $-2\pm20$ \\
    HD~61005  &$85.9\pm0.1$ & $70.4\pm0.1$&$8\pm22$ & $-4\pm8$ \\
    HD~92945  &$65.4\pm0.6$ & $100.0\pm0.6$&$-84\pm55$ & $12\pm35$ \\
    HD~107146 &$19.3\pm0.6$ & $153.2\pm1.7$&$36\pm35$ & $-48\pm35$\\
    HD~109573 &$76.6\pm0.1$ & $26.5\pm0.0$ &$11\pm2$ & $-32\pm2$ \\
    HD~121617 &$44.1\pm0.6$ & $58.7\pm0.7$&$8\pm8$ & $14\pm7$ \\
    \hline
    HD~84870  &$47.1\pm2.6$ & $-25.7\pm4.1$&$81\pm57$ & $-103\pm63$ \\
    HD~95086  &$31.6\pm3.5$ & $100.0\pm5.7$&$-91\pm45$ & $127\pm34$ \\
    HD~131488 &$85.0\pm0.1$ & $97.2\pm0.0$&$0\pm2$ & $1\pm1$ \\
    HD~131835 &$74.2\pm0.2$ & $59.2\pm0.2$ &$6\pm6$ & $3\pm6$ \\
    HD~218396 &$28.8\pm3.3$ & $49.7\pm5.7$&$-41\pm65$ & $-57\pm69$\\
    \hline
    \end{tabular}
    \label{tab:sourceproperties}
    \tablefoot{The parameters defined here all match those of ARKS~I, Table~B.1; see \cite{overview_arks}.}
\end{table*}

\section{Offset systems}
\label{app:checkSources}

\begin{table*}
    \centering
    \caption{Offset analysis table.}
    \begin{tabular}{l|cc|cc|c}
    \hline
    \hline
    Science Target & Check Source & Number of EBs & {\tt imfit} offset & {\tt uvmodelfit} offset & Disc centre offset\\
     &  & & $\Delta$RA, $\Delta$Decl. [mas] & $\Delta$RA, $\Delta$Decl, [mas] & $\Delta$RA, $\Delta$Decl. [mas] \\
    \hline
    HD~15115  & J0239+0416 & 11 & $1.6{\pm}6.7$, $3.3{\pm}6.4$ & $0.7{\pm}5.8$, $-3.3{\pm}5.8$ & $44{\pm}12$, $-8{\pm}5$\\    
    HD~32297  & J0519+0848 & 4  & $0.5{\pm}3.9$, $-0.5{\pm}4.8$ & $-0.7{\pm}1.7$, $1.7{\pm}2.6$ & $19{\pm}4$, $18{\pm}4$\\
    HD~109573 & J1254-4424 & 2  & $-2.4{\pm}0.8$, $5.9{\pm}2.4$ & $1.3{\pm}0.7$, $-5.4{\pm}2.6$ & $11{\pm}2$, $-32{\pm}2$\\
    \hline
    \end{tabular}
    \label{tab:checksources}
    \tablefoot{This table compiles the image and uv-plane analysis metrics for check source offsets, their number of execution blocks (EBs) for the three science targets, and the best-fit disc centres \citep[from ARKS~I;][]{overview_arks} for the sources determined to have significant offsets between those positions and their stellar locations.}
\end{table*}
ALMA's astrometric accuracy depends on a number of factors, including the atmospheric phase conditions, the separation between the science target and the phase calibrator(s), errors in antenna positions, and the delay model used during data correlation (see ALMA Technical Handbook, Chapter 10.5.2, Astrometric Observations).
The astrometric accuracy for a given observation can be estimated empirically by measuring the offset position of a well-known calibrator across multiple observations.
In the case of these `check sources', these have locations that are known with extremely high precision, and are typically ICRF quasars with positions determined by Very Long Baseline Interferometry (VLBI) observations.
The a-priori positions of these ICRF quasars are accurate to ${<}0.0001''$, thus based on the baseline configurations during ARKS ALMA observations, the check source phases should have offsets consistent with zero.
In other words, if the stellocentric offsets observed towards any ARKS targets, though most pertinent to this appendix, HD~15115, HD~32297 and HD~109573, were solely due to phase calibration uncertainties, the same offsets would be present in the data for the check sources.

Check source selection from the ALMA Calibrator Source Catalogue is based on criteria that ensures check sources are at comparable distances from the phase calibrator as the science target, and that the check source must be sufficiently bright to achieve a S/N greater than 15 across all scans and spectral windows. 
The check source should also maintain adequate strength to support reliable imaging during both the QA0+ analysis and the ALMA Pipeline data processing stages.
Research on long baseline observations from 2017 indicates that the peak-to-total flux density ratio serves as a dependable metric for evaluating observation quality. 
This metric is influenced by the root-mean-square (rms) phase difference between phase calibrator scans and the angular separation from the phase calibrator (refer to ALMA Technical Handbook, Sect. 10.4.10).

We used the check sources present in the observations of HD~109573, HD~15115, and HD~32297 to assess the positional accuracy of the calibrated data. 
In all cases, the same check source was observed for all execution blocks (EBs) per-science target, allowing for a consistent comparison of astrometric accuracy between different observations. 
Comparison between the offset position in different days and observing conditions provides an estimate of the absolute astrometric precision of ALMA.
We measured the check source location for all execution blocks, and in both the image and visibility planes.
Specifically, we imaged the check source after averaging all scans using the same {\tt CASA tclean} parameters as were used to image the science target, and fitted a 2D Gaussian to the images with {\tt CASA's imfit} tool.
We compared the centroid coordinates of this Gaussian fit to the phase centre of the check source observation, corresponding to the ICRS/J2000 VLBI position from the ALMA Calibrator Source Catalogue.
Likewise, we modelled the check source visibilities as point sources in {\tt CASA's uvmodelfit} tool, and made the same comparison between the fitted centroid location and that of the check source.
Check sources are expected to have an absolute positional errors ${<}0.002''$ and appear as point sources at ALMA’s resolution (though decoherence may cause artificial source extension). 
To estimate the astrometric accuracy, we then calculated the repeatability of the measured positions, defined as the dispersion in the measured check source positions across all execution blocks for the three check sources. 
The resulting mean offset and dispersion for all execution blocks (per target or check source) are presented in Table~\ref{tab:checksources}.
For both RA and Decl., the first value represents the mean, while the quoted error corresponds to the dispersion of all measurements.

For HD~109573, two observations were conducted using J1254-4424 as the check source. 
The science target and the check source are located $4.99^\circ$ and $4.90^\circ$ from the phase calibrator, respectively. 
Given their similar distances from the phase calibrator, the astrometric accuracy measured for the check source serves as a reliable proxy for the accuracy toward HD~109573.
For HD~32297, four observations were conducted using J0519+0848 as the check source. 
The science target and the check source are located $2.47^\circ$ and $3.49^\circ$ from the phase calibrator, respectively. 
Since the target is closer to the phase calibrator than the check source, the astrometric accuracy measured for the check source can be considered an upper limit for the target.
Finally, for HD~15115, 12 observations were conducted using J0239+0416 as the check source. 
In this case, the science target and the check source are located $0.82^\circ$ and $4.69^\circ$ from the phase calibrator, respectively. 
Since the phase RMS is proportional to $\sqrt{d}$, with $d$ being the geometric distance between the lines of sight of the target and the phase calibrator \citep[see][and references therein]{Maud+2020,Maud+2022}, the estimated astrometric accuracy toward HD~15115 is $\approx 1/\sqrt{{\rm diff}}$ of the accuracy estimated for the check source, i.e. given the angular difference of $4.69^\circ-0.82^\circ=3.87^\circ$, a factor of $1/\sqrt{3.87} \approx 1/2$.
We note that one observation was excluded from the analysis of HD~15115's data to a phase RMS close to $60^\circ$, approximately four times higher than the average phase RMS of the other observations. 
Thus, 11 observations were included in total. 

Both the mean offset and scatter (standard deviation) are of the order of a few milliarcseconds, a factor of 5--10 below ALMA's predicted astrometric accuracy \citep{Lestrade2008}, but consistent with phase referencing experiments \citep[carried out during long baselines campaigns; these experiments indicate that ALMA's measured absolute calibrator positions have a scatter of a few milliarcseconds; see][]{ALMAPart15}.

\section{Self-subtraction maps}
\label{sec:app_selfsubmaps}
\begin{figure*}
    \centering
    \includegraphics[clip, trim={0cm 2.0cm 1cm 2.5cm}, width=1.0\linewidth]{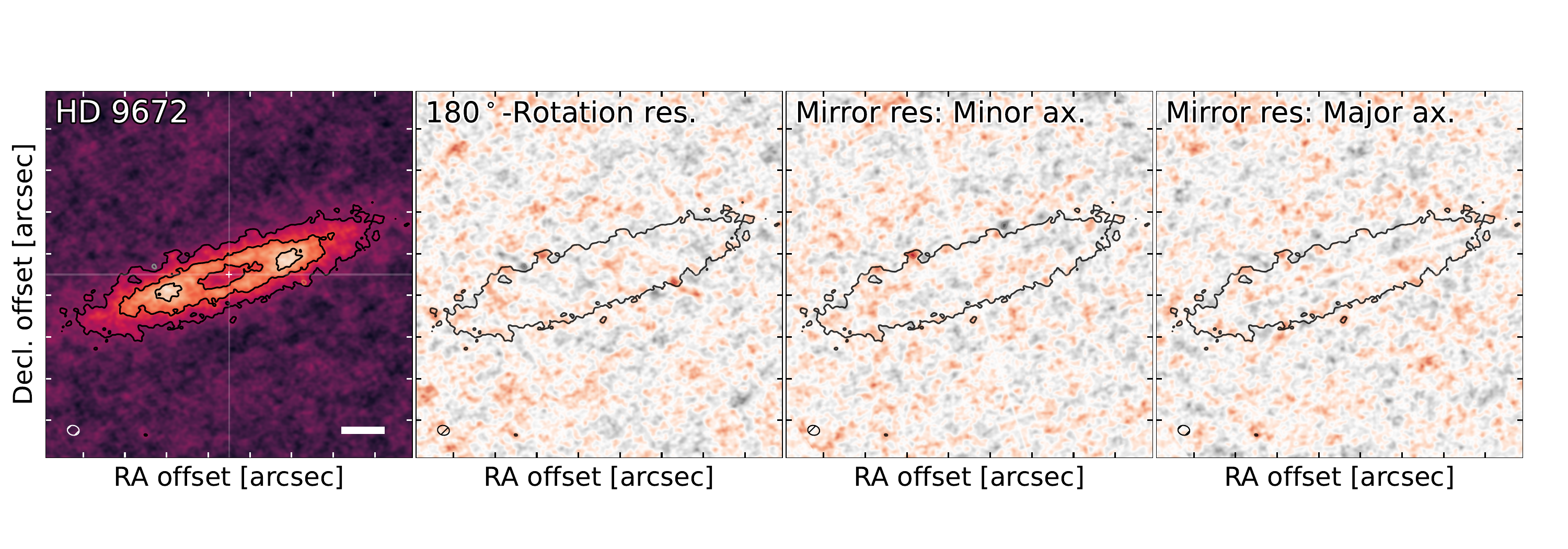}
    \includegraphics[clip, trim={0cm 2.0cm 1cm 2.5cm}, width=1.0\linewidth]{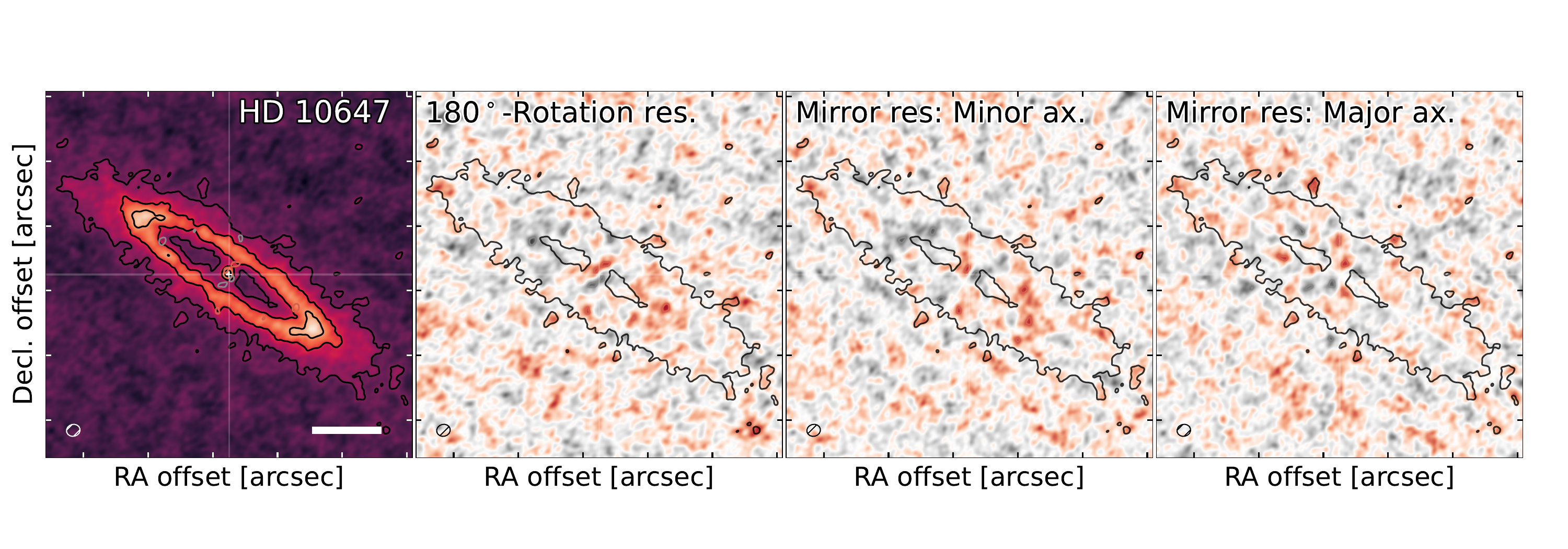}
    \includegraphics[clip, trim={0cm 2.0cm 1cm 2.5cm}, width=1.0\linewidth]{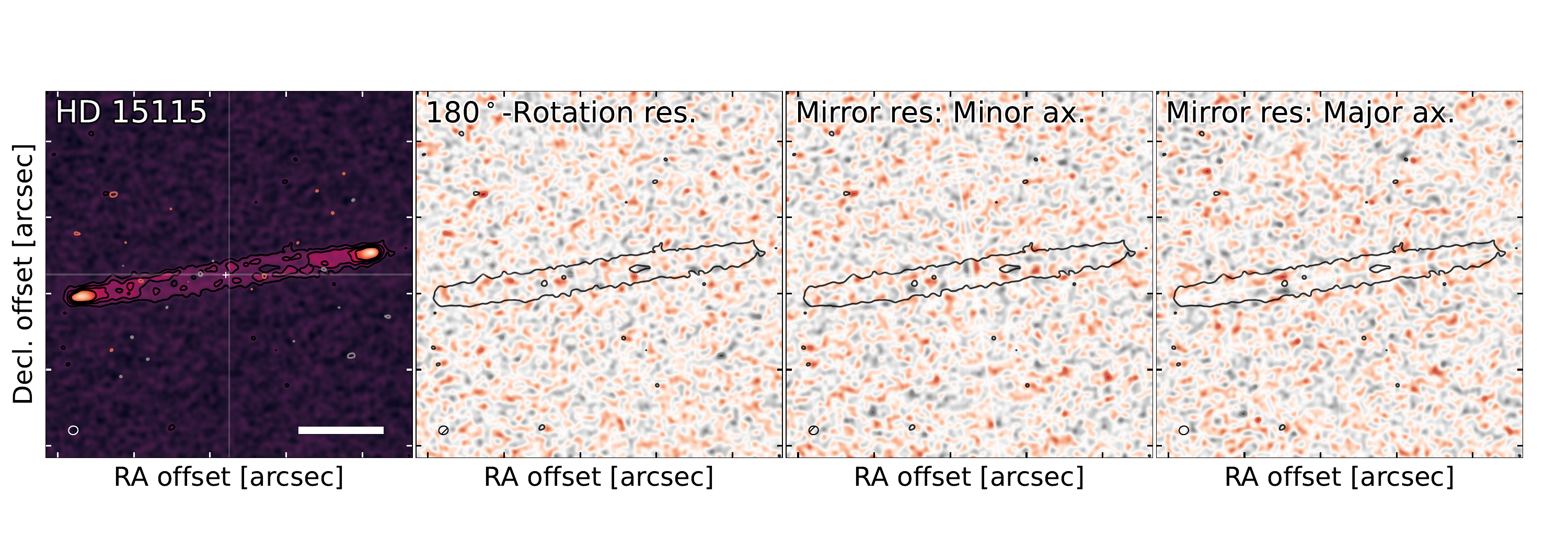}
    \includegraphics[clip, trim={0cm 2.0cm 1cm 2.5cm}, width=1.0\linewidth]{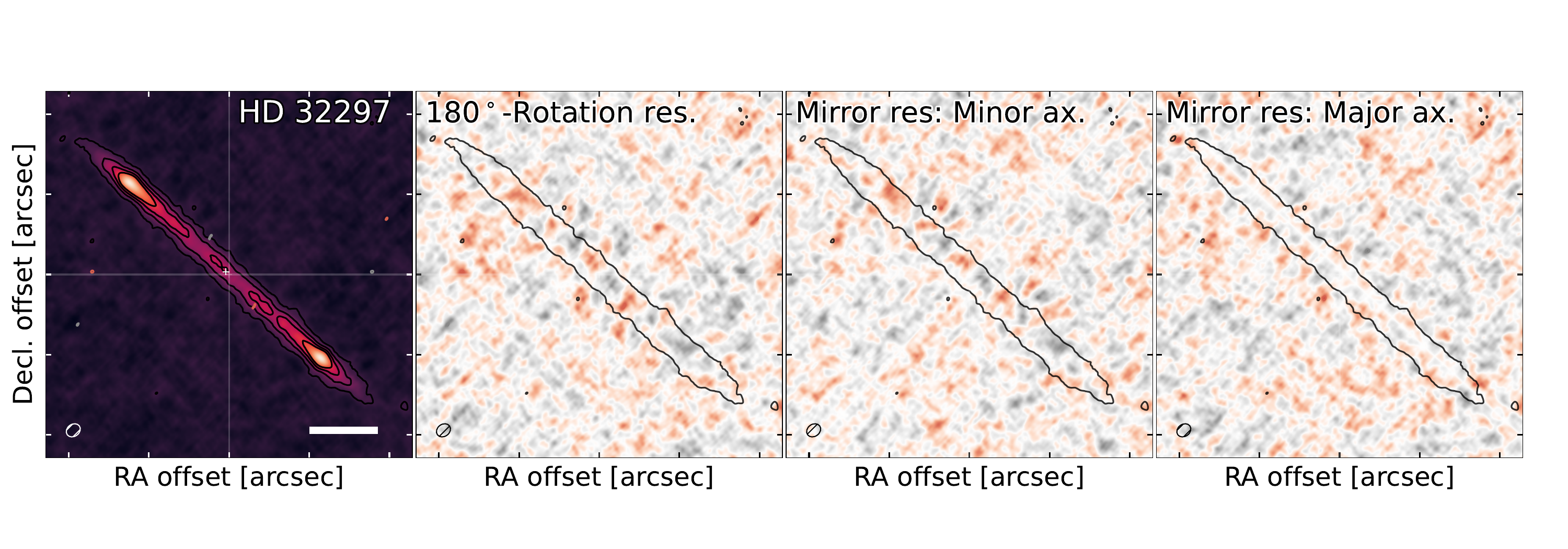}
    \caption{Self-subtraction residual maps for HD~9672, HD~10647, HD~15115, and HD~32297. The plots show their images (left), rotation-subtraction residuals (centre left), and mirror-subtractions along the minor and major axes (centre right and right, respectively).  Residual maps show the $3\sigma$ image emission contour (black), and ${\pm}3\sqrt{2}\times\sigma$ and $5\sqrt{2}\times\sigma$ residual emission contours (grey and red respectively for negative and positive). 
    The colour scale ranges from ${\pm}5\sigma$ of the rms in the residual maps.
    Beams are shown in the lower left of each panel, and 50\,au scale bars are shown on the lower right of the left panel. 
    We present the primary beam-corrected contours of the rotation self-subtraction residuals on the left panel to highlight where these arise from in the image plane for HD~9672, HD~15115, and HD~32297, and from the minor axis self-subtraction residuals for HD~10647.}
    \label{fig:selfsubgallery1}
\end{figure*}

\begin{figure*}
    \centering
    \includegraphics[clip, trim={0cm 2.0cm 1cm 2.5cm}, width=1.0\linewidth]{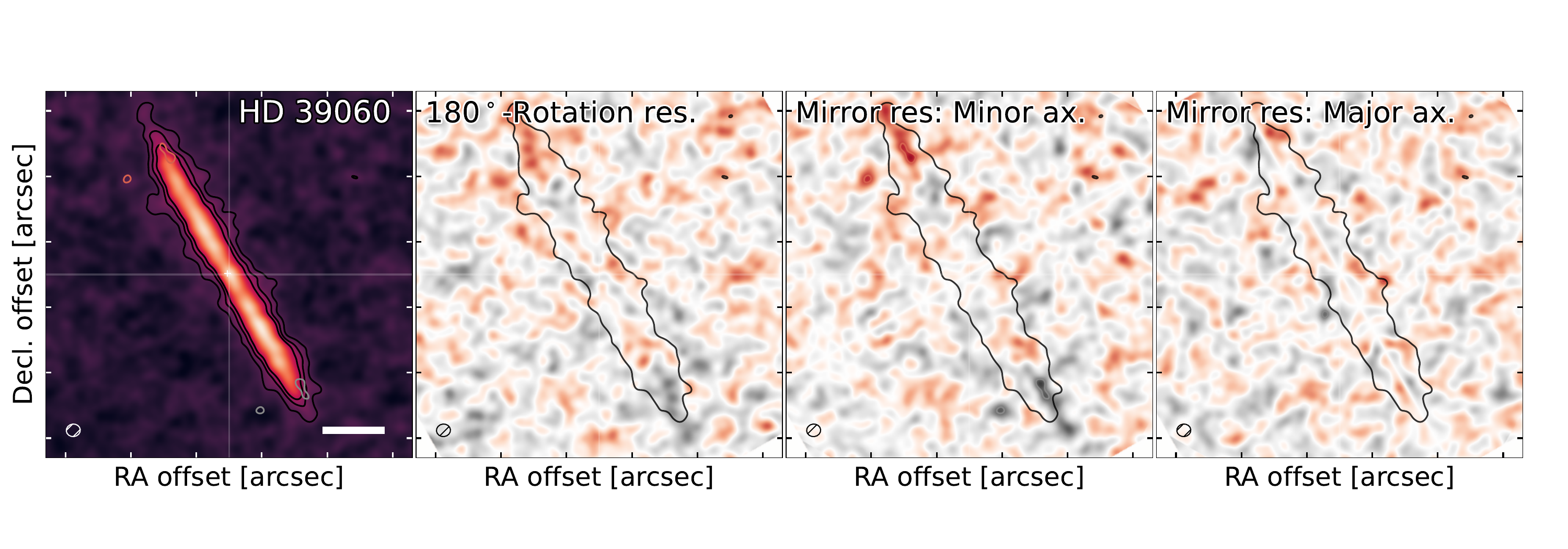}
    \includegraphics[clip, trim={0cm 2cm 1cm 2.5cm}, width=1.0\linewidth]{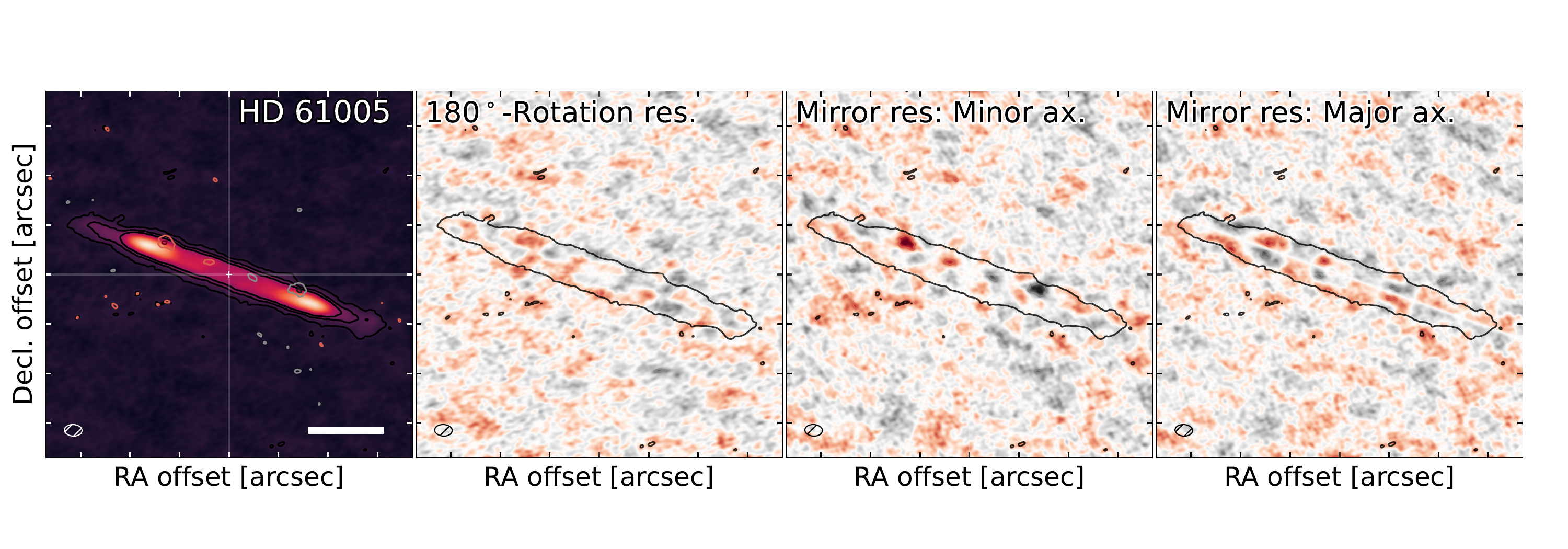}
    \includegraphics[clip, trim={0cm 2.0cm 1cm 2.5cm}, width=1.0\linewidth]{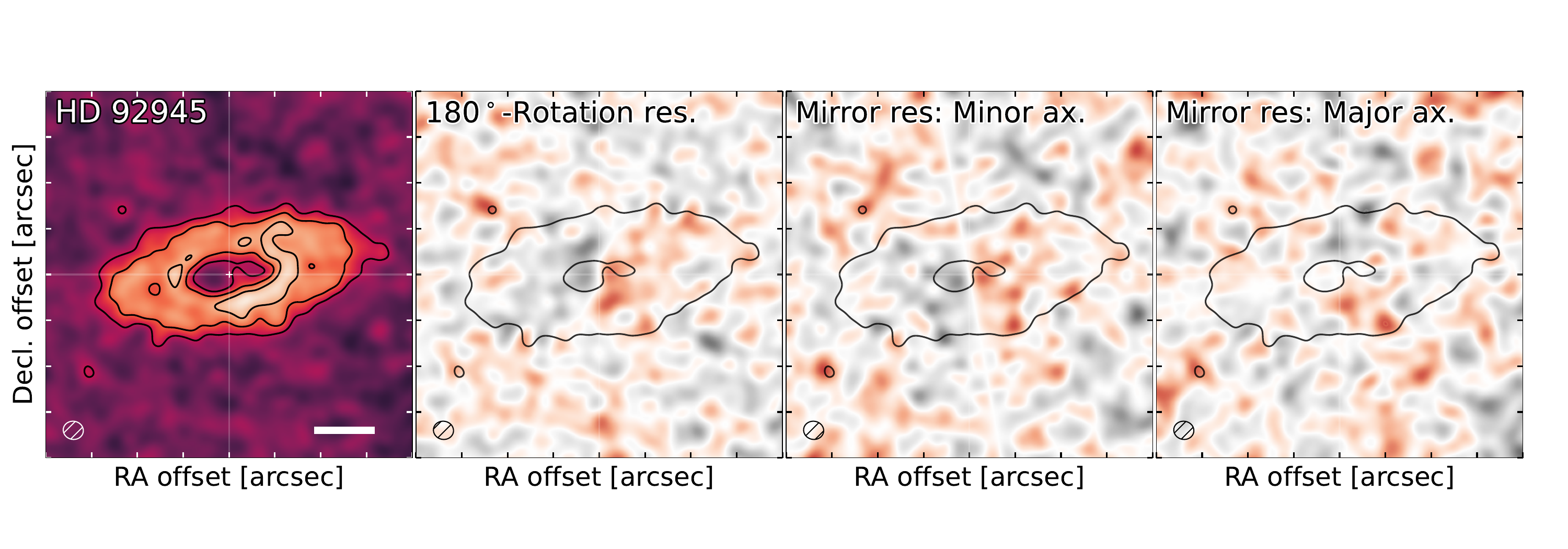}
    \includegraphics[clip, trim={0cm 2cm 1cm 2.5cm}, width=1.0\linewidth]{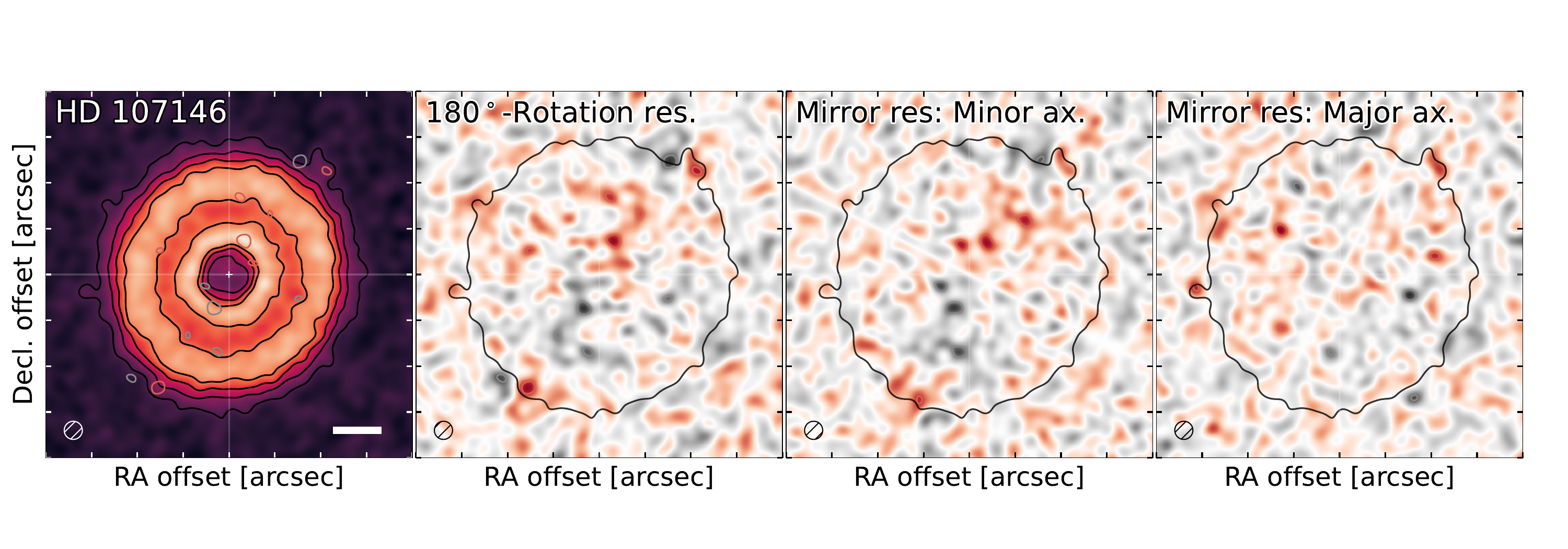}
    \caption{Self-subtraction residual maps for HD~39060, HD~61005, HD~92945, and HD~107146. This figure presents data in the same manner as Fig.~\ref{fig:selfsubgallery1}. We present the contours of the rotation self-subtraction residuals in the left panels to highlight where these arise from in the image plane for HD~92945 and HD~107146, and the minor axis self-subtraction residuals for HD~39060 and HD~61005.}
    \label{fig:selfsubgallery2}
\end{figure*}

\begin{figure*}
    \centering
    \includegraphics[clip, trim={0cm 2.0cm 1cm 2.5cm}, width=1.0\linewidth]{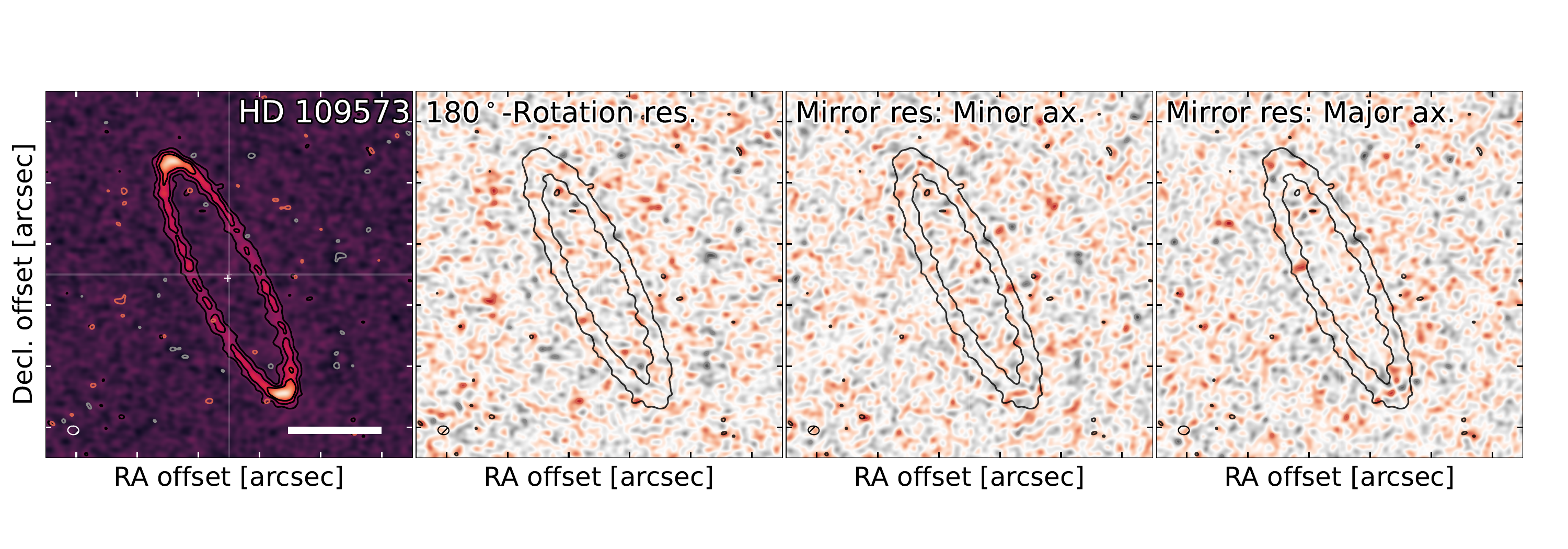}
    \includegraphics[clip, trim={0cm 2cm 1cm 2.5cm}, width=1.0\linewidth]{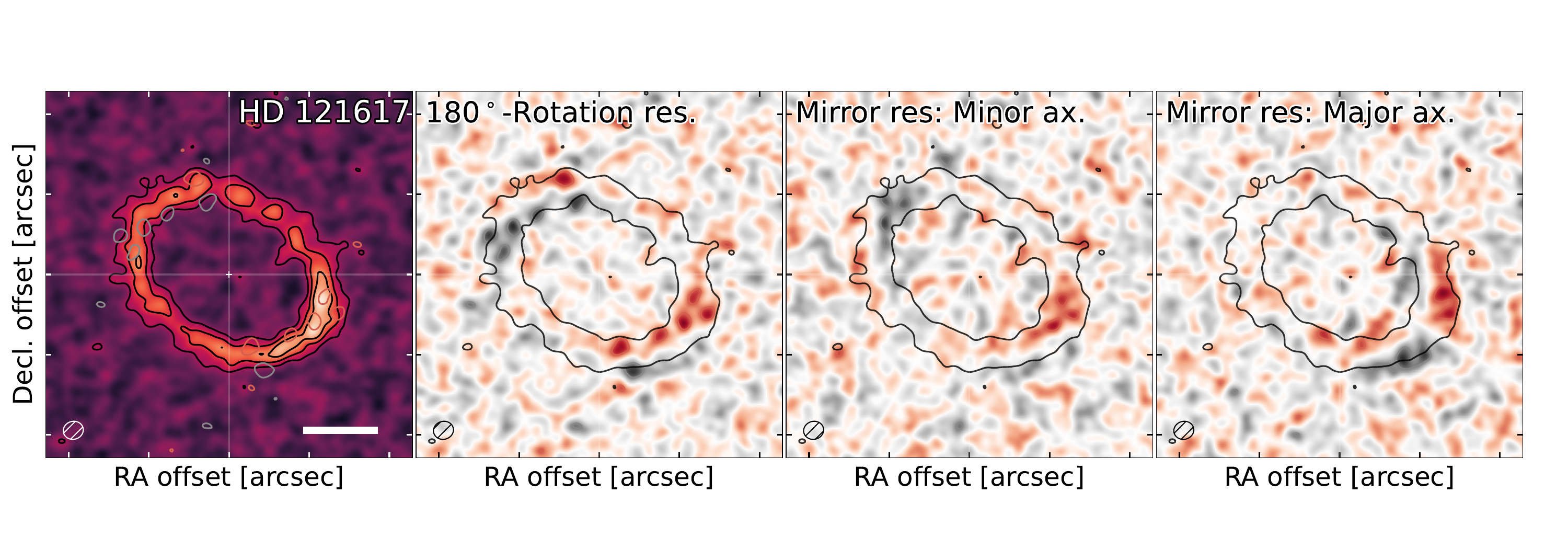}    
    \caption{Self-subtraction residual maps for HD~109573 and HD~121617. This figure presents data in the same manner as Fig.~\ref{fig:selfsubgallery1}. We present the contours of the rotation self-subtraction residuals in the left panels to highlight where these arise in the image plane for both sources.}\label{fig:selfsubgallery3}
\end{figure*}

\begin{figure*}
    \centering
    \includegraphics[clip, trim={0cm 2.0cm 1cm 2.5cm}, width=1.0\linewidth]{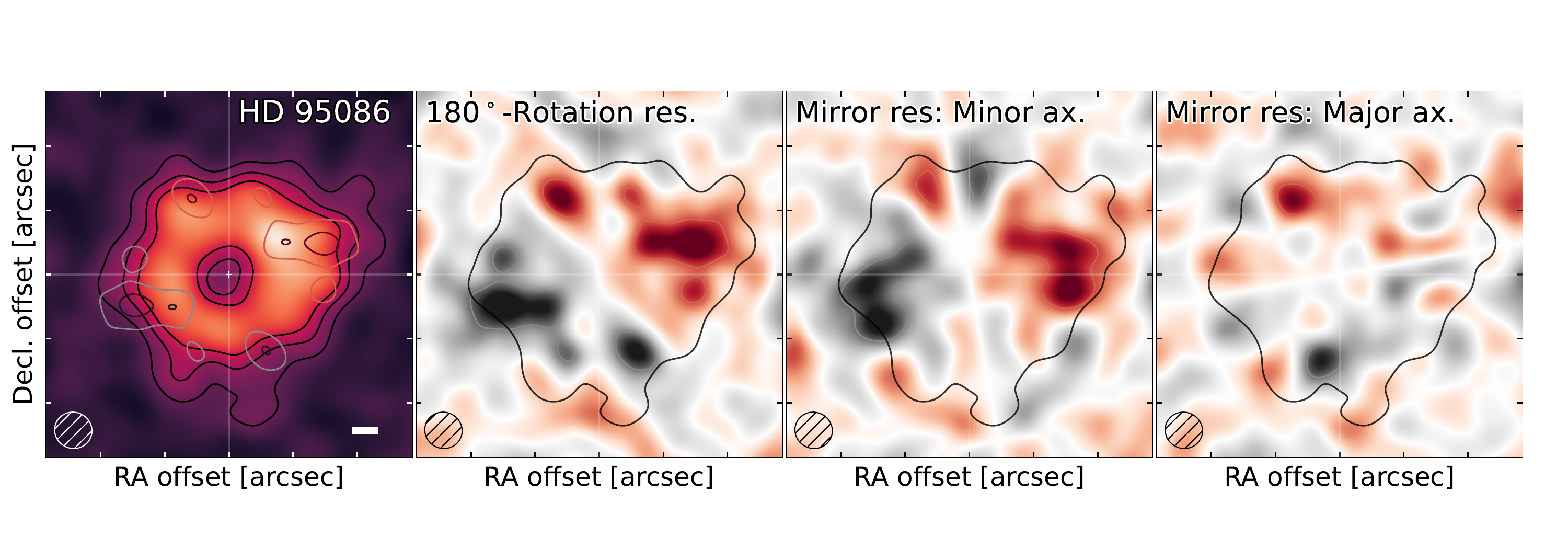}
    \caption{Self-subtraction residual maps for HD~95086, which due to undersubtraction of a background galaxy is host to a strong asymmetry that is unlikely to be physically located in the disc. This figure presents data in the same manner as Fig.~\ref{fig:selfsubgallery1}.}
\label{fig:selfsubgallery4}
\end{figure*}

\begin{figure*}
    \centering
    \includegraphics[clip, trim={0cm 2.0cm 1cm 2.5cm}, width=1.0\linewidth]{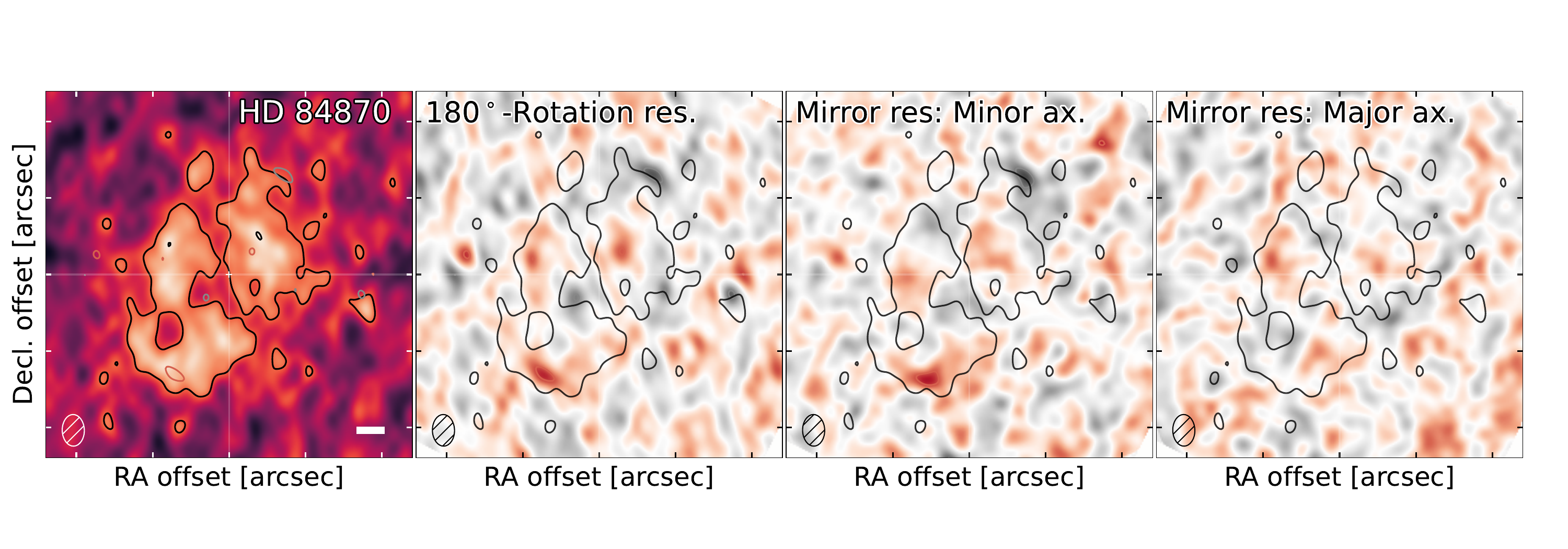}
    \includegraphics[clip, trim={0cm 2.0cm 1cm 2.5cm}, width=1.0\linewidth]{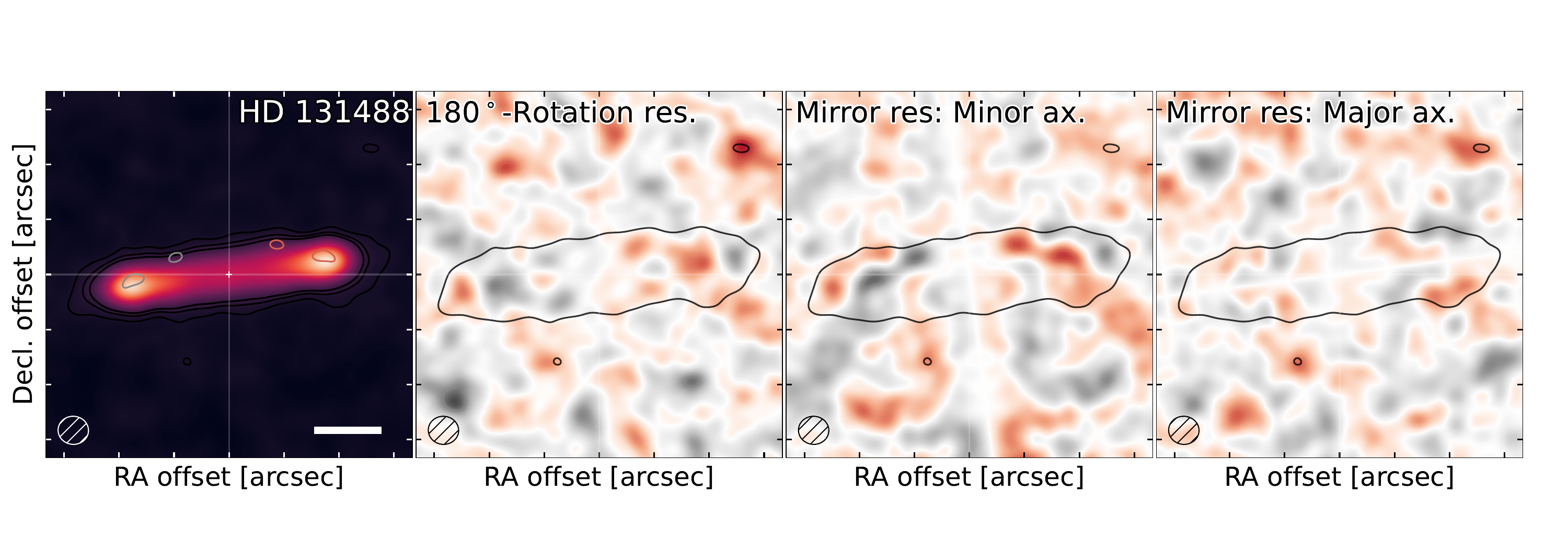}
    \includegraphics[clip, trim={0cm 2cm 1cm 2.5cm}, width=1.0\linewidth]{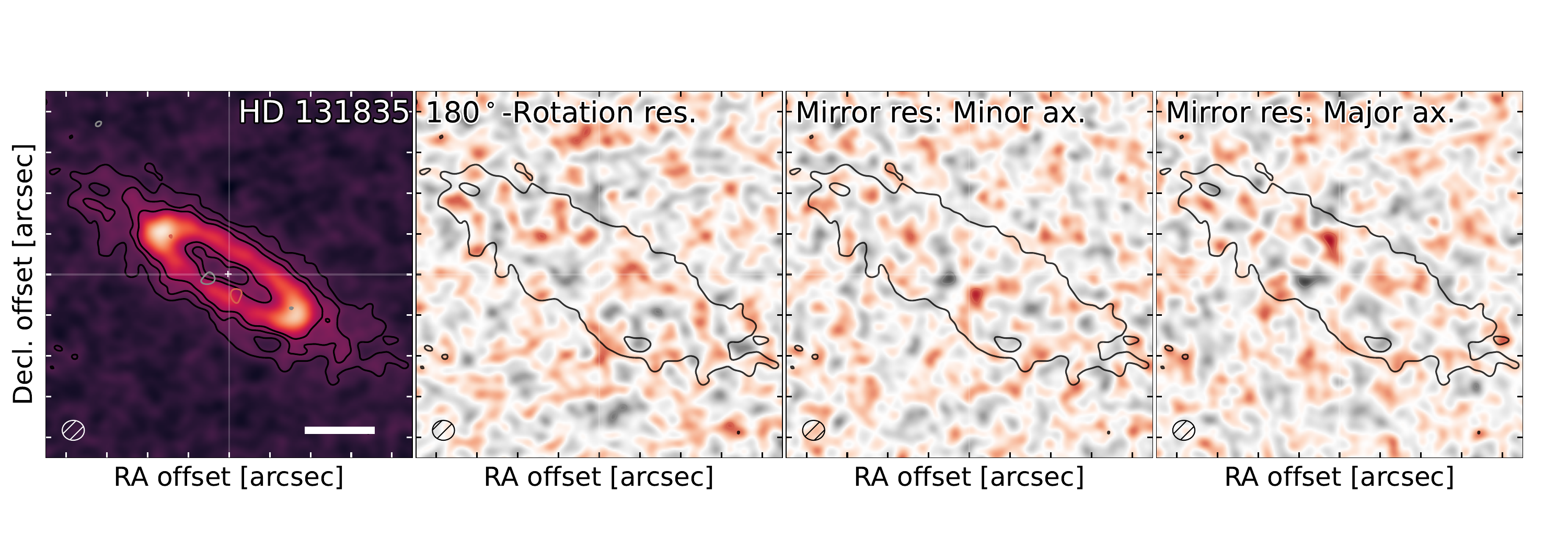}
    \includegraphics[clip, trim={0cm 2cm 1cm 2.5cm}, width=1.0\linewidth]{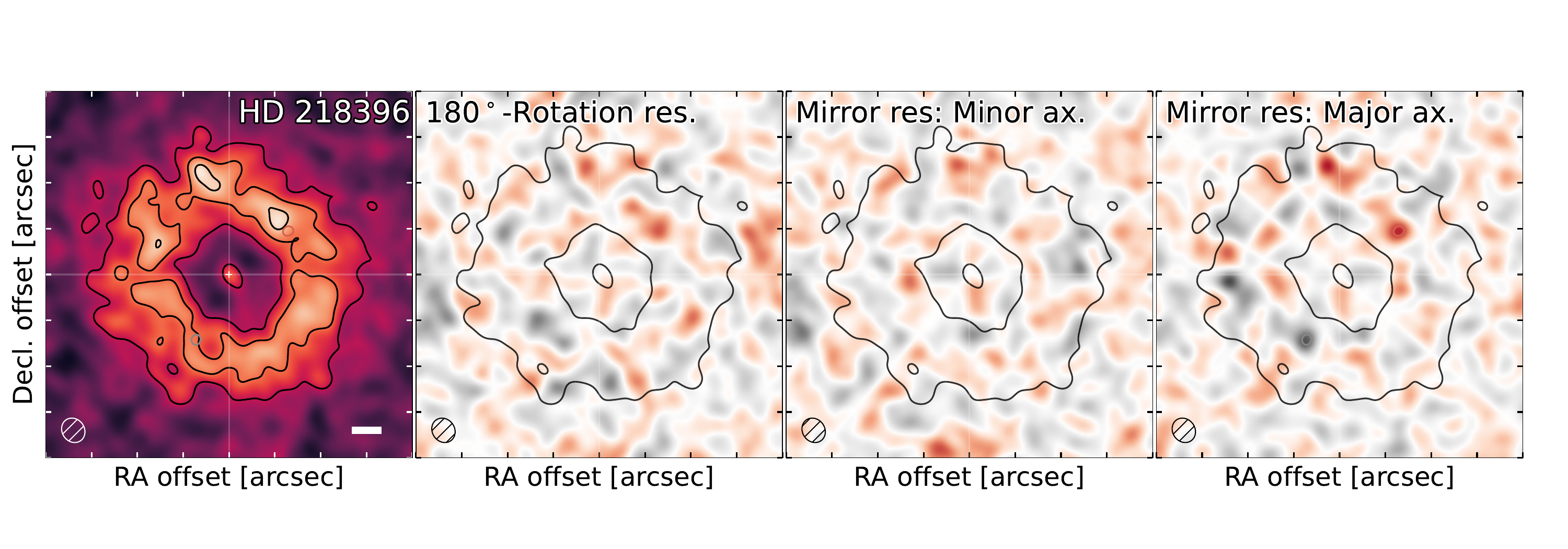}
    \caption{Self-subtraction residual maps for systems with tentative signs of asymmetries (HD~84870, HD~131488, HD~131835, and HD~218396). This figure presents data in the same manner as Fig.~\ref{fig:selfsubgallery1}.  We present the contours of the rotation self-subtraction residuals in the left panel to highlight where these arise from in the image plane for HD~84870, of the minor axis self-subtraction residuals for HD~131488 and HD~131835, and of the major axis self-subtraction residuals for HD~218396.}
    \label{fig:selfsubgallery55}
\end{figure*}

\begin{figure*}
    \centering
    \includegraphics[clip, trim={0cm 0cm 0cm 0cm}, width=1.0\linewidth]{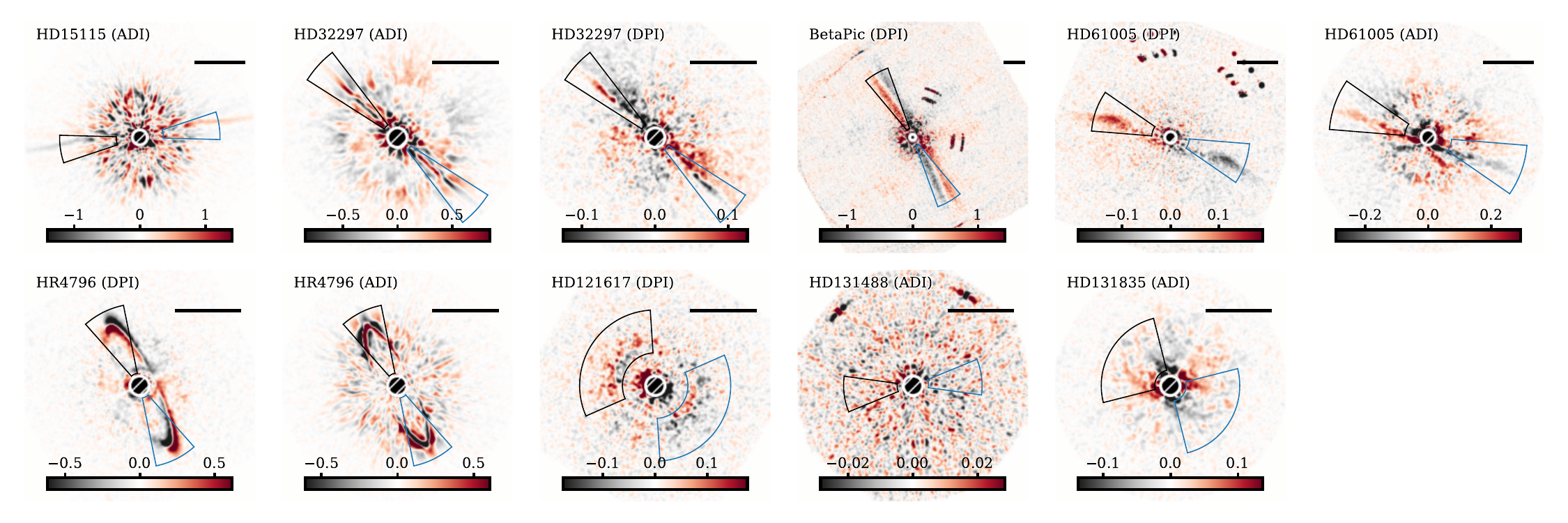}
    \includegraphics[clip, trim={0cm 0cm 0cm 0cm}, width=1.0\linewidth]{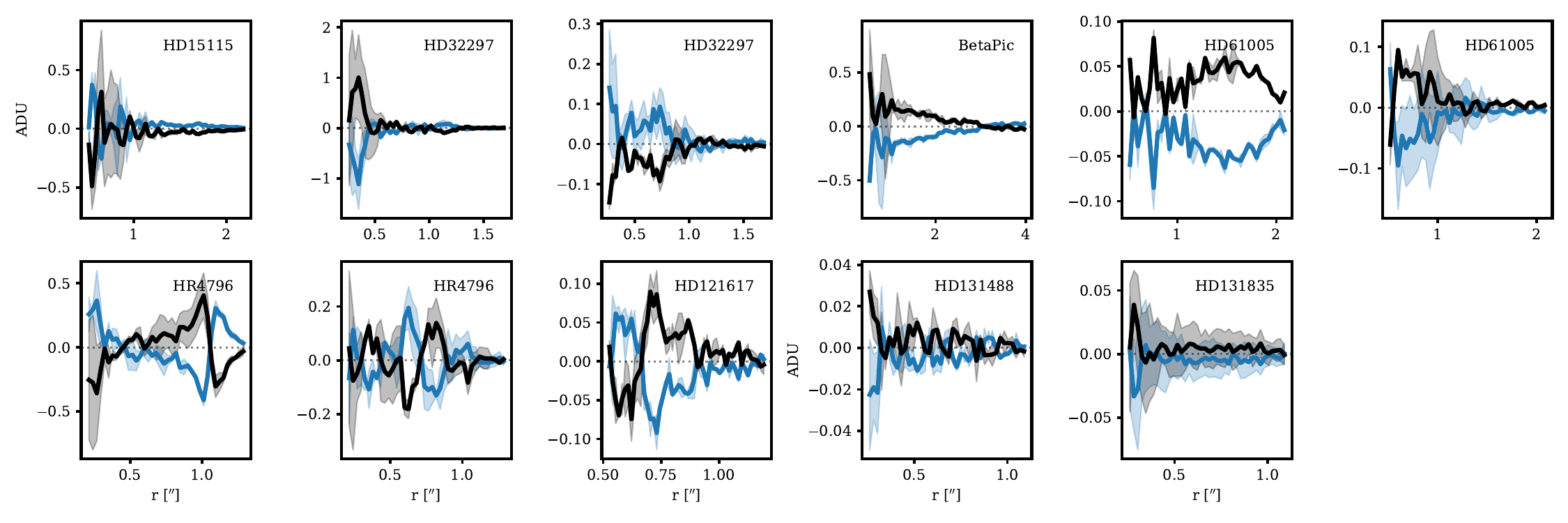}    
    \caption{Top: Minor-axis self-subtraction residual maps for systems with SPHERE-detected emission. The black bar in the top right corner of each panel indicates $1$\arcsec, denoted as either angular differential imaging (ADI) or dual polarisation imaging (DPI), depending on whether the total intensity maps or the Stokes $Q_\phi$ component from \citet{scat_arks} are used. Bottom: Radial profiles through the self-subtracted residual maps, estimated inside the wedges shown in the top panel for each corresponding image. The colour-coding of the wedges matches the colour-coding of the radial profiles (black and blue). The profiles present estimated ${\pm}1\sigma$ uncertainties.}
    \label{fig:selfsubgallery5}
\end{figure*}

Here we present galleries of the self-subtraction maps of sources host to asymmetric emission or tentative asymmetries in Figs.~\ref{fig:selfsubgallery1}, ~\ref{fig:selfsubgallery2}, ~\ref{fig:selfsubgallery3}, ~\ref{fig:selfsubgallery4}, and ~\ref{fig:selfsubgallery55}. 
For the systems HD~9672, HD~10647, HD~39060, HD~61005, HD~92945, HD~107146, HD~121617, HD~95086, HD~84870, and HD~218396 we used image centres based on the Gaia DR3 stellar location, whereas for the systems HD~15115, HD~32297, HD~109573, HD~131488 and HD~131835 we used image centres based on the best-fit disc centres derived in \citet{overview_arks}; see also Table~\ref{tab:sourceproperties}.
In 
Fig.~\ref{fig:selfsubgallery5} we also present a gallery of SPHERE self-subtracted (minor axis) maps and their associated radial profiles, showing the extent to which these sources are asymmetric in scattered light.

\end{appendix} 
\end{document}